\documentclass[aps,prx,twocolumn,showpacs,superscriptaddress,amsmath,amssymb,groupedaddress]{revtex4-1}

\usepackage{graphicx}
\usepackage[colorlinks=true,linkcolor=blue,citecolor=blue,urlcolor=blue]{hyperref}
\usepackage{bbold}
\usepackage{gensymb}
\usepackage{multirow}
\usepackage{xcolor}
\usepackage{colortbl}
\usepackage{hhline}

\makeatletter

    \def\CT@@do@color{%
      \global\let\CT@do@color\relax
            \@tempdima\wd\z@
            \advance\@tempdima\@tempdimb
            \advance\@tempdima\@tempdimc
    \advance\@tempdimb\tabcolsep
    \advance\@tempdimc\tabcolsep
    \advance\@tempdima2\tabcolsep
            \kern-\@tempdimb
            \leaders\vrule
                    \hskip\@tempdima\@plus  1fill
            \kern-\@tempdimc
            \hskip-\wd\z@ \@plus -1fill }
    \makeatother
    
\AtBeginDocument{%
    \newwrite\bibnotes
    \def\bibnotesext{Notes.bib}
    \immediate\openout\bibnotes=\jobname\bibnotesext
    \immediate\write\bibnotes{@CONTROL{REVTEX41Control}}
    \immediate\write\bibnotes{@CONTROL{%
    apsrev41Control,author="08",editor="1",pages="0",title="0",year="1"}}
     \if@filesw
     \immediate\write\@auxout{\string\citation{apsrev41Control}}%
    \fi
}%

\def \a {\alpha}

\def \d {\delta}
\def \D {\Delta}
\def \e {\epsilon}
\def \ve {\varepsilon}

\def \g {\gamma}

\def \l {\lambda}
\def \L {\Lambda}
\def \o {\omega}

\def \s {\sigma}

\def \dag {\dagger}

\def \p {\partial}

\def \apx {\approx}

\def \til {\tilde}

\def \dag {\dagger}

\newcommand{\sumv}[1]{\sum_{\mbf #1}}

\newcommand{\sumvdp}[1]{\sum_{\mbf #1\mbf #1'}}

\def \rar {\rightarrow}

\def \la {\langle}
\def \ra {\rangle}
\def \fr {\frac}
\def \lf {\left}
\def \ri {\right}

\newcommand{\ket}[1]{|#1\ra}
\newcommand{\braket}[3]{\la#1|#2|#3\ra}
\newcommand{\innp}[2]{\la#1|#2\ra}

\newcommand{\epvl}[1]{\la#1\ra}

\def \Tr {\mathrm{Tr}}
\def \bece {\begin{center}}
\def \ence {\end{center}}
\def \beeq {\begin{equation}}
\def \eneq {\end{equation}}
\def \beal {\begin{aligned}}
\def \enal {\end{aligned}}
\def \bega {\begin{gathered}}
\def \enga {\end{gathered}}
\def \benu {\begin{enumerate}}
\def \ennu {\end{enumerate}}
\def \beit {\begin{itemize}}
\def \enit {\end{itemize}}
\def \bede {\begin{description}}
\def \ende {\end{description}}
\def \betb {\begin{tabular}}
\def \entb {\end{tabular}}
\def \bear {\begin{array}}
\def \enar {\end{array}}

\def \mbf {\mathbf}

\def \mca {\mathcal}

\def \txt {\text}

\newcommand{\comment}[1]{}

\begin{document}


\title{Chiral flat band superconductivity from symmetry-protected three-band crossings}

\author{Yu-Ping Lin}
\affiliation{Department of Physics, University of Colorado, Boulder, Colorado 80309, USA}

\date{\today}

\begin{abstract}
We show that chiral (nearly) flat band superconductivity can develop and host novel Majorana fermions at a time-reversal pair of symmetry-protected three-band crossing points. Based on symmetry analysis, mean-field study, and superfluid stiffness calculation, we determine and analyze the irreducible pairing channels with flat band pairings in the low-energy spin-$1$ fermion theory. Flat band pairing can enhance superconductivity dramatically, where the critical temperature scales linearly in the interaction strength. While fully gapped flat band pairing states develop in the single-component pairing channels, we find chiral $\bar p\pm i\bar p$ flat band superconductivity in the multicomponent pairing channels. Three-dimensional itinerant Majorana fermions arise at the bulk nodal points, whereas Majorana arcs appear on the surface.
\end{abstract}

\maketitle

\section{Introduction}

Chiral superconductivity has attracted much attention of modern condensed matter research in the past decades \cite{kallin16rpp,sato17rpp}. Hosting finite angular momentum pairing, chiral superconductivity spontaneously breaks time-reversal symmetry and manifests nontrivial topological properties. Two-dimensional (2D) chiral superconductivity exhibits an out-of-plane rotation axis, thereby manifesting a fully gapped quasiparticle spectrum in the bulk \cite{read00prb}. Such a state has been studied and proposed extensively in various systems, including strontium-based materials \cite{mackenzie03rmp,fischer14prb}, graphene systems \cite{nandkishore12np,lin19prb}, and fractional quantum Hall states \cite{read00prb,son18anncmp}. Meanwhile, three-dimensional (3D) chiral superconductivity manifests various types of gap structures. The bulk can host either full gaps, nodal points, nodal lines, or nodal Fermi surfaces, depending on whichever band structure and symmetry are provided \cite{sigrist91rmp,nandkishore16prb,agterberg17prl,brydon18prb,venderbos18prx}. The best-known example of a 3D chiral pairing state is the superfluid $^3$He-$A$ phase \cite{volovikbook}. Recent works have also proposed 3D chiral superconductivity in various other materials, such as heavy fermion compounds \cite{joynt02rmp} and topological semimetals \cite{nandkishore16prb,brydon16prl,agterberg17prl,boettcher18prl,venderbos18prx,brydon18prb,roy99prb}.

Another mainstream of modern condensed matter research has focused on superconductivity with high ratio of critical temperature over Fermi temperature $T_c/T_F$. Two different classes of systems have been uncovered along this direction, where the comparison of bandwidth $W$ and interaction $V$ plays a crucial role. The first class is represented by the high-$T_c$ materials \cite{lee06rmp}, where strong electronic interaction $V\gg W$ induces high critical temperature below which superconductivity develops. The other class manifests low-energy bands with (nearly) flat dispersion, $W\ll V$. Remarkably, the pairing critical temperature acquires a {\it linear} scaling $T_c^\txt{pair}\sim V$ in the {\it flat band} limit $W\rar0$, owing to the immense density of states \cite{miyahara07pc,kopnin11prb,heikkila11jetp,uchoa13prl,lin18prb}. Although the obstruction to global phase coherence is usually expected in flat band pairing, {\it anomalous phase coherence} may still develop on flat bands in multiband systems \cite{peotta15nc,julku16prl,liang17prb,xie20prl}. This phase coherence is captured by an anomalous part of superfluid stiffness, which indicates a phase coherence critical temperature again with linear scaling $T_c^\txt{phase}\sim V$. The true critical temperature of superconductivity is then determined by the stronger fluctuation $T_c=\min\{T_c^\txt{pair},T_c^\txt{phase}\}$, with the linear scaling $T_c\sim V$ generally manifested. Such dramatic enhancement has motivated an intensive search for flat band superconductivity in practical materials. Various 2D systems have been studied accordingly, including surfaces of gapless topological materials \cite{heikkila11jetp}, strained graphene \cite{uchoa13prl}, and graphene moir\'e heterostructures \cite{cao18n,chen19n}.

Motivated by these mainstreams of modern condensed matter research, we consider a platform where chiral superconductivity may develop on the 3D flat bands. Our analysis addresses the pairing problem at the {\it three-band crossing points}. Three-band crossings have been theoretically proposed under various conditions \cite{bradlyn16sc,fulga17prb,zhu17pra,hu18prl,ramires19prb}, while experimental realization has been accomplished in the superconducting quantum circuit systems \cite{tan18prl}. Here we focus on a time-reversal pair of symmetry-protected three-band crossing points. These band crossings may occur at the high-symmetry points in 3D time-reversal-symmetric materials with nonsymmorphic space-group symmetries \cite{bradlyn16sc}. Notably, the $\mbf k\cdot\mbf p$ Hamiltonian at these points manifests the novel {\it spin-1 fermions}. Due to the opposite exchange properties in the spin sector, these fermions exhibit significant difference from spin-$1/2$ electrons in the pairing problem. Moreover, the low-energy theory manifests a middle flat band at the band crossing, hosting an immense density of states. The pairing on flat bands can support dramatically enhanced superconductivity, with the critical temperature scaling linearly in the interaction strength \cite{lin18prb}.

In this paper, we further confirm that chiral superconductivity can benefit from these flat bands. Based on symmetry analysis, mean-field study, and superfluid stiffness calculation, we examine the irreducible pairing channels with various valley pairings and spin-orbit coupled pairings with total angular momenta $J=0,1$. Our analysis focuses on the channels with flat band pairings, which raise the critical temperatures to the linear scaling of interest. Previous analysis has studied the single-component $J=0$ pairing channels, which manifest fully gapped quasiparticle spectra \cite{lin18prb}. Here we find {\it chiral} $\bar p\pm i\bar p$ {\it flat band superconductivity} in the multicomponent $J=1$ pairing channels. These chiral pairing states are the spin-nondegenerate analogy of the superfluid $^3$He-$A$ phase \cite{volovikbook}. The bulk nodal points arise in the quasiparticle spectra and host {\it 3D itinerant Majorana fermions} \cite{venderbos16prb,kozii16sa}, remarkably different from the Weyl points in the superfluid $^3$He-$A$ phase. Open {\it Majorana arcs} are also uncovered on the surface accordingly \cite{meng12prb,sau12prb,venderbos16prb,kozii16sa}. We thus uncover a dramatically enhanced 3D chiral flat band superconductivity where novel Majorana fermions can arise.

\section{Low-energy theory}

\begin{figure}[b]
\centering
\includegraphics[scale = 1]{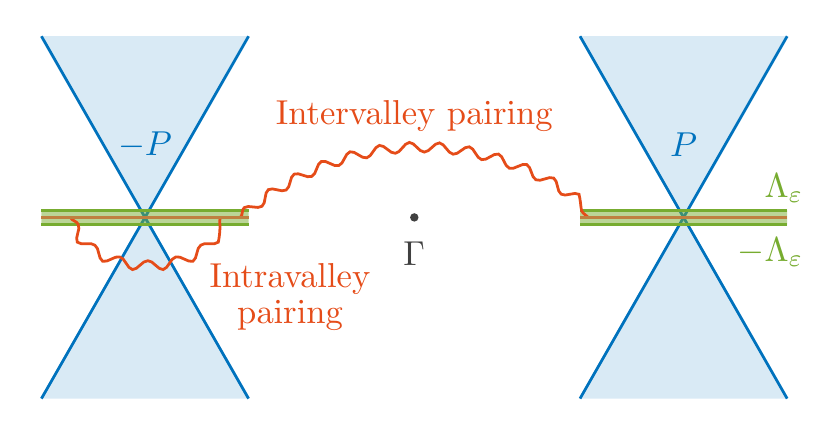}
\caption{\label{fig:3bc} Illustration of symmetry-protected three-band crossing points. Near the band crossings $\mu\apx0$, pairings on the flat band (brown lines) are dominant in superconductivity. The linear bands are projected out from the pairing since the attractive regime $[\mu-\L_\ve,\mu+\L_\ve]$ is narrow.}
\end{figure}

In time-reversal symmetric materials with space group 199, a pair of three-band crossings is stabilized at the high-symmetry points $\pm P$ with momenta $\mbf P_\pm=\pm\mbf P$ (Fig.~\ref{fig:3bc}) \cite{bradlyn16sc}. Time-reversal symmetry enforces these band crossings to occur at the same energy. The low-energy behavior is described by an effective two-valley spin-$1$ fermion theory
\beeq
\label{eq:ham0}
H_0=\sum_{\l=\pm}\sumv{k}c_{\l\mbf k}^\dag\mca H_{0\mbf k}c_{\l\mbf k},\quad
\mca H_{0\mbf k}=v\mbf k\cdot\mbf S-\mu,
\eneq
where the minimal $\mbf k\cdot\mbf p$ Hamiltonian $\mca H_{0\mbf k}$ is exhibited in the vicinity of $\pm P$. Here $\mu$ is the chemical potential, $v$ is the effective velocity, and $\mbf k$ is the relative momentum from $\pm P$ with cutoff $k<\L_k$. The fermion field $c_{\mbf k}^\dag$ contains six valley-spin indices $(\l=\pm,s=1,0,-1)$. According representations are formed by Pauli matrices $\l^{0,1,2,3}$ and spin-$1$ operators $S^i$'s in the $S^z$ eigenbasis. The low-energy theory is invariant under time reversal $\mca T=(i\l^2)\g K$, which swaps the valleys, inverts the momentum, and flips the spin. Here $\g=\exp(i\pi S^y)$ and $K$ is the complex conjugate operator. An approximate rotation symmetry around $\pm P$ is also manifest.

At each band crossing $\pm P$, the eigenstates describe the three bands with energies $\ve_{\mbf k}^{0,\pm}-\mu$. Two of the bands are linear, $\ve_{\mbf k}^\pm=\pm vk$, and the middle band is flat, $\ve_{\mbf k}^0=0$. The wavefunctions manifest monopole harmonics $\ket{u_{\mbf k}^a}=\sqrt{4\pi/3}(Y_{q^a11}^*,Y_{q^a10}^*,Y_{q^a1-1}^*)^T$ with monopole charges $q^\pm=\mp1$ and $q^0=0$ \cite{wu76npb,wu77prd}. Significantly, the monopole charge $q^a$ corresponds to the Chern number $C^a=2q^a$ from the Berry flux calculation \cite{xiao10rmp}. This implies that the band crossings are topologically nontrivial, with $C^\pm=\mp2$ on the nontrivial linear bands and $C^0=0$ on the trivial flat band. Time-reversal symmetry imposes the same Chern numbers at $\pm P$. Nonzero net Chern numbers are compensated by additional Weyl points in the Brillouin zone, which can be away from the low-energy regime.

\section{Flat band superconductivity}

We wish to study potential superconductivity in the vicinity of band crossings. In particular, we focus on the pairing states with dramatic enhancement from flat band pairings. Despite their various possible origins, the pairing states can be classified and studied based on symmetry, Fermi statistics, and topology \cite{cho12prb,savary17prb,venderbos18prx,lin18prb}. Here we identify the irreducible pairing channels based on a symmetry analysis, then study the flat band pairing states that can arise in these channels.

\subsection{Irreducible pairing channels}

Based on the symmetry of low-energy theory (\ref{eq:ham0}), the pairings can be distinguished into various irreducible pairing channels. Each channel manifests a pairing operator $c_{\mbf k}^\dag\mca M[(i\l^2)\g(c_{-\mbf k}^\dag)^T]$. The pairing representation $\mca M$ is labeled by a set of good quantum numbers under the symmetry. Note that parity is not a good quantum number since there is no inversion symmetry.

Since the model manifests valley $\txt{SU}(2)_\txt{v}$ symmetry, the valley pairings can be distinguished into singlet $\a=0$ and triplet $\a=3,\pm$ channels. These channels manifest pairing representations $\l^{0,3}/\sqrt2$ and $\l^\pm=(\l^1\pm i\l^2)/2$. Similar to the pairing of spin-$1/2$ states, the valley pairing is antisymmetric and symmetric in singlet $\a=0$ and triplet $\a=3,\pm$ channels, respectively. Note that different pairing channels exhibit different pairing momenta (Fig.~\ref{fig:3bc}) \cite{cho12prb}. The intervalley pairings $\a=0,3$ manifest zero momentum and lead to Bardeen-Cooper-Schrieffer (BCS) states. Meanwhile, the intravalley pairings $\a=\pm$ carry finite momenta $\pm2\mbf P$, thereby triggering pair density wave (PDW) states instead \cite{agterberg20ar}. One may also consider the combination of PDW states $\a=1,2$ in order to have time-reversal symmetry.

Rotation symmetry further distinguishes the pairings into different angular momentum channels. Spin-orbit coupling enforces the good quantum numbers $(L,S,J,M_J)$, where $L$ is orbital angular momentum, $S$ is spin, and $J=L+S$ is the total angular momentum with axial component $M_J$. The pairings of spin-$1$ fermions are remarkably different from those of ordinary spin-$1/2$ electrons. With the larger single-particle spin, more spin modes $S=0,1,2$ are available under pairing. More importantly, the pairings show opposite exchange properties to the conventional ones. While spin singlet and quintet pairings $S=0,2$ are symmetric, spin triplet pairing $S=1$ is antisymmetric. This important difference can lead to novel pairing states which are absent in spin-$1/2$ systems.

The combination of valley and spin-orbit coupled pairings is constrained by Fermi statistics. With valley singlet pairing $\a=0$, the spin-orbit coupled pairing should be even. The even-$L$ states must carry even $S$, while the odd-$L$ states should come with odd $S$. When the valley pairing is triplet $\a=0,\pm$, the spin-orbit coupled pairing should be odd. This swaps the coupling between $L$ and $S$ quantum numbers, with even-$L$ states carrying odd $S$ and vice versa. These combinations form the classification of irreducible pairing channels in the low-energy theory (\ref{eq:ham0}). Note that the opposite exchange properties in the spin pairing have led to different combinations from those in the spin-$1/2$ electronic systems. For example, valley triplet even-$L$ pairing states exhibit spin singlet pairing $S=0$ in Weyl semimetals \cite{cho12prb}, while spin triplet pairing $S=1$ is necessary for the spin-$1$ fermions herein.

Projected on a single irreducible pairing channel, the interaction takes the form \cite{savary17prb,venderbos18prx,lin18prb}
\beeq
\label{eq:int0}
H_\txt{int}=-\fr{V}{\mca V}\sumvdp{k}(\vec P_{J\mbf k})^\dag\cdot\vec P_{J\mbf k'},
\eneq
where $\mca V$ denotes the spatial volume of the system. Each channel manifests a pairing operator $(\vec P_{J\mbf k})^\dag=c_{\mbf k\l_+}^\dag\vec J_{J\mbf k}[\g(c_{-\mbf k\l_-}^\dag)^T]$, where the valley indices $\l_\pm$ are determined by the valley pairing channel $\a$. The $2J+1$ irreducible representations $J_{JM_J\mbf k}=\sum_{M_LM_S}\innp{LS;M_LM_S}{JM_J}L_{LM_L\mbf k}S_{SM_S}$ characterize the spin-orbit coupled pairings $(L,S,J,M_J)$, where the additions are determined by the Clebsch-Gordan coefficients $\innp{LS;M_LM_S}{JM_J}$. The orbital representations $L_{LM_L\mbf k}=\sqrt{4\pi}(k/\L_k)^LY_{LM_L\hat{\mbf k}}$ describe the $2L+1$ orbital modes $(L,M_L)$ and manifest spherical harmonics. Meanwhile, the spin representations $S_{SM_S}\sim Y_{SM_S}(\mbf S)$ (with anticommutators involved) are $3\times3$ $\txt{SU}(2)$ irreducible representations with normalization $\Tr(S_{SM_S}S_{S'M_{S'}'}^\dag)=\d_{SS'}\d_{M_SM_{S'}'}$. They describe the $2S+1$ spin pairings $(S,M_S)$ of spin-$1$ fermions. Constant attraction $-V<0$ is assumed in each channel, which is a proper setup for our general study of potential superconductivity. The constant assumption is eligible for short-range interactions. Whether a channel is attractive depends on the mechanism inducing superconductivity.

\subsection{Mean-field theory}

When superconductivity develops in a channel, the system acquires a finite condensate of according pairing. Such a pair condensate is captured by a finite order parameter $\vec\D(T)=-(V/\mca V)\sumv{k}\epvl{\vec P_{J\mbf k}}_T$, with $\epvl{\cdots}_T$ denoting the ensemble average at temperature $T$. The mean-field Hamiltonian takes the Bogoliubov-de Gennes (BdG) form
\beeq
H=\sumv{k}\Psi_{\mbf k}^\dag\mca H_{\txt{BdG},\mbf k}\Psi_{\mbf k}.
\eneq
Here the BdG Hamiltonian reads
\beeq
\label{eq:bdg3}
\mca H_{\txt{BdG},\mbf k}=\lf(\bear{cc}\mca H_{0\mbf k}&\vec\D\cdot\vec J_{J\mbf k}\\\vec{\bar\D}\cdot\vec J_{J\mbf k}^\dag&-\mca H_{0\mbf k}\enar\ri),
\eneq
and $\Psi_{\mbf k}^\dag=(c_{\mbf k\l_+}^\dag,[\g(c_{-\mbf k\l_-}^\dag)^T]^\dag)$ represents the Nambu spinor. The eigenstates are referred to as BdG quasiparticles. Due to an intrinsic particle-hole symmetry of the BdG Hamiltonian, these quasiparticles come in particle-hole pairs $b$ with opposite energies $\pm E^b_{\mbf k}$. Various properties of superconductivity can be uncovered by solving the self-consistent gap equation \cite{lin18prb}
\beeq
\label{eq:gapeq3}
\bar\D_{M_J}=\sum_b\fr{V}{\mca V}\sumv{k}\fr{\p E^b_{\mbf k}}{\p\D_{M_J}}\tanh\fr{E^b_{\mbf k}}{2T}.
\eneq
These include the pairing critical temperature $T_c^\txt{pair}$ below which Cooper pairs develop, and the order parameter magnitude $|\vec\D(T)|$, as well.

\subsection{Effective flat band theory}

Our interest lies in the potential superconductivity in the vicinity of band crossings $\mu\apx0$. The middle flat band is dominant in this regime, as it manifests a divergent density of states $\nu^0(\ve)=n^0\d(\ve)$ where $n^0$ represents the total number of flat band states per unit volume. Meanwhile, the linear bands are irrelevant since their densities of states are vanishing, $\nu^\pm(\ve)\rar0$ as $\ve\rar0$. Generically, the mechanisms inducing superconductivity only manifest a narrow attractive window $|\ve-\mu|<\L_\ve\ll v\L_k$ near the Fermi level (Fig.~\ref{fig:3bc}). This excludes almost the whole linear bands and leaves only the flat band for pairing. Whether superconductivity develops thus depends solely on the flat band pairing.

\subsubsection{BdG theory}

With a focus on the flat band pairing, the effective Hamiltonian $H^0=\sumv{k}(\Psi^0_{\mbf k})^\dag\mca H^0_{\txt{BdG},\mbf k}\Psi^0_{\mbf k}$ is obtained via a direct projection on flat bands. The Nambu spinor now takes the form $(\Psi^0_{\mbf k})^\dag=([c^0_{\mbf k\l_+}]^\dag,c^0_{-\mbf k\l_-})$, and the BdG Hamiltonian (\ref{eq:bdg3}) becomes
\beeq
\label{eq:bdg0}
\mca H^0_{\txt{BdG},\mbf k}=\lf(\bear{cc}\ve^0_{\mbf k}-\mu&\D_{\mbf k}\\\D_{\mbf k}^*&-(\ve^0_{\mbf k}-\mu)\enar\ri).
\eneq
Note that the gap function on the flat bands $\D_{\mbf k}=\vec\D\cdot\vec J^0_{J\mbf k}$ has been defined, where the pairing representation $\vec J^0_{J\mbf k}=\braket{u^0_{\mbf k}}{\vec J_{J\mbf k}}{u^0_{\mbf k}}$ is a vector of $2J+1$ $\mbf k$-dependent scalars. The quasiparticle energies $\pm E^0_{\mbf k}=\pm[(\ve^0_{\mbf k}-\mu)^2+|\D_{\mbf k}|^2]^{1/2}$ can be solved directly from the BdG Hamiltonian.

In the effective flat band theory, the gap equation (\ref{eq:gapeq3}) reduces to
\beeq
\label{eq:gapeq0}
1=\fr{V}{\mca V}\sumv{k}\fr{|J^0_{JM_J\mbf k}|^2}{2E^0_{\mbf k}}\tanh\fr{E^0_{\mbf k}}{2T}.
\eneq
We solve this gap equation with the chemical potential fixed at band crossing $\mu=0$. The solution shows that the characteristic energy scales of superconductivity, namely, pairing critical temperature and zero-temperature order parameter magnitude, are linear in the interaction strength
\beeq
\label{eq:linscal}
T_c^\txt{pair},|\vec\D(0)|\sim Vn^0
\eneq
on similar order \cite{lin18prb}. In a conventional BCS state, these energy scales are usually exponentially small $T_c^\txt{pair},|\vec\D(0)|\sim\exp[-1/V\nu(\mu)]$, as the density of states $\nu(\mu)$ on a normal Fermi surface is finite. The linear scaling here indicates that pairing is dramatically enhanced at three-band crossings $\pm P$. Such enhancement is possible solely because the whole flat bands contribute divergent densities of states to the pairing.

\subsubsection{Anomalous phase coherence}

With the pairing dramatically enhanced on flat bands, it is worth discussing whether superconductivity can develop at temperatures as high as $T_c^\txt{pair}$. Although such an expectation may seem plausible at first sight, the crucial role of phase fluctuations in flat bands can lead to a significantly different conclusion \cite{peotta15nc}. For single-band systems, Cooper pairs become well localized and lose the communication with one another in the flat band limit. With the global phase coherence obstructed, Cooper pairs on flat bands cannot form superconductivity where the Meissner effect and dissipationless supercurrent are manifest \cite{tinkhambook}. Nevertheless, `anomalous phase coherence' can develop on flat bands when multiband structure (multiorbital in real space) is involved \cite{peotta15nc,julku16prl,liang17prb,xie20prl}. In certain multiband systems, the exponentially localized Wannier orbitals may only be achieved by mapping a set of bands onto real space. Examples include the systems where topologically nontrivial bands are present. Accordingly, a single flat band may map onto broadened Wannier orbitals which overlap with one another. Although the normal states are nondispersive, the Cooper pairs can still develop global phase coherence through the overlaps. Therefore, flat band superconductivity may find potential realization in these multiband systems, including the three-band crossing system of our interest.

The anomalous phase coherence is intimately related to the multiorbital geometry on a flat band. This feature is indicated by an anomalous part of superfluid stiffness \cite{peotta15nc,julku16prl,liang17prb,xie20prl}. In the effective flat band theory (\ref{eq:bdg0}), the anomalous superfluid stiffness takes the form
\beeq
\label{eq:sfstf}
D^s_{\txt{geom},ij}(T)=\fr{1}{\mca V}\sumv{k}\fr{2|\D_{\mbf k}|^2}{E^0_{\mbf k}}\tanh\fr{E^0_{\mbf k}}{2T}g_{ij\mbf k},
\eneq
which measures the response of supercurrent $j^s_i$ to the external gauge field $A_j$. The involvement of the quantum metric $g_{ij\mbf k}=(1/2)(\innp{\p_{k_i}u^0_{\mbf k}}{\p_{k_j}u^0_{\mbf k}}+\innp{\p_{k_j}u^0_{\mbf k}}{\p_{k_i}u^0_{\mbf k}})+\innp{u^0_{\mbf k}}{\p_{k_i}u^0_{\mbf k}}\innp{u^0_{\mbf k}}{\p_{k_j}u^0_{\mbf k}}$ indicates the contribution from band geometry, as it captures the variation of the flat band eigenstate $d^2(\mbf k,\mbf k+d\mbf k)=1-|\innp{u^0_{\mbf k}}{u^0_{\mbf k+d\mbf k}}|^2=g_{ij\mbf k}dk_idk_j$ in momentum space. When the flat band eigenstate acquires a $\mbf k$ dependence in the composition of orbitals, according Wannier orbitals become broadened in real space and overlap with one another. Such behavior is precisely described by a finite quantum metric $g_{ij\mbf k}$, which leads to a finite anomalous superfluid stiffness $D^s_{\txt{geom},ij}(T)$. According to these features, the anomalous superfluid stiffness $D^s_{\txt{geom},ij}(T)$ indeed captures the anomalous phase coherence which may develop in flat band pairing.

The components of the quantum metric for the flat band eigenstate $\ket{u^0_{\mbf k}}$ read $g_{ii\mbf k}=(k^2-k_i^2)/k^4$ and $g_{ij\mbf k}=-k_ik_j/k^4$ for $i\neq j$. Fixing the chemical potential at band crossing, $\mu=0$, we find that the anomalous superfluid stiffness is proportional to the order parameter magnitude $D^s_{\txt{geom},ij}(T)\sim|\vec\D(T)|\L_k$. In particular, the diagonal components $D^s_{\txt{geom},ii}(0)$ at zero temperature are on similar order to $|\vec\D(0)|\L_k$, while the off-diagonal components with $i\neq j$ are usually subleading. As the simplest example, we present the results in the pairing channel $(L,S,J)=(0,0,0)$ with constant gap function $\D_{\mbf k}=\D$. The superfluid stiffness is diagonal in this channel, with the isotropic diagonal components $D^s_{\txt{geom},ii}(0)=(2/3\pi^2)|\D(0)|\L_k$.

The critical temperature for phase coherence, $T_c^\txt{phase}$, can be obtained from the superfluid stiffness. While a rigorous result in two dimensions can be derived from the Berezinskii-Kosterlitz-Thouless (BKT) theory \cite{kosterlitz73jpc}, a general solution in three dimensions has not been available \cite{hazra19prx}. Nevertheless, an estimation based on a 3D $XY$ model can be made following Emery and Kivelson $T_c^\txt{phase}\sim[\prod_iD^s_{\txt{geom},ii}(0)]^{1/3}\xi$ \cite{emery95n}, where the zero-temperature superfluid stiffness is adopted. Based on Pippard's argument with the uncertainty principle \cite{tinkhambook}, we further estimate the coherence length by the reciprocal of available momentum range, $\xi\sim\L_k^{-1}$ \cite{engelbrecht97prb,hazra19prx}. Remarkably, these estimations yield a linear scaling for the phase coherence critical temperature
\beeq
T_c^\txt{phase}\sim Vn^0,
\eneq
which is on similar order to the pairing critical temperature: $T_c^\txt{phase}\sim T_c^\txt{pair}$. The true critical temperature of superconductivity is then determined from the competition between pair and phase fluctuations,
\beeq
\label{eq:tc}
T_c=\min\{T_c^\txt{pair},T_c^\txt{phase}\}\sim Vn^0,
\eneq
where the stronger fluctuation dominates. Note that the linear scaling in interaction strength is always valid for the critical temperature $T_c$. Despite the potential obstruction by phase fluctuations, flat band superconductivity can still develop at a dramatically high critical temperature at the three-band crossing (\ref{eq:ham0}).

\begin{figure}[b]
\centering
\includegraphics[scale = 1]{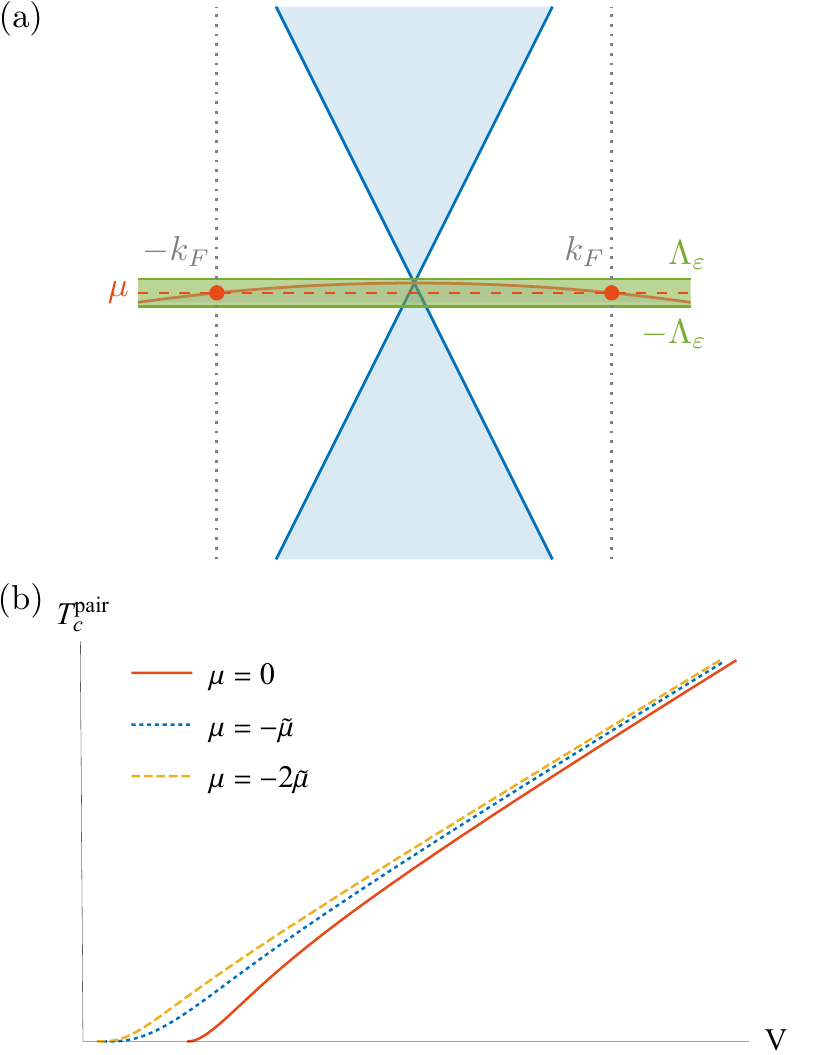}
\caption{\label{fig:bandcurv} Band curvature effect. (a) Quadratic perturbation results in an infinitesimal band curvature on the flat band (perturbation on linear bands is safely neglected in this illustration). The Fermi level is set at an infinitesimal doping $\mu<0$. At the Fermi momenta, the large interband gap projects out the multiband pairing. (b) Scaling of pairing critical temperature $T_c^\txt{pair}$ in the interaction strength $V$. Here $-\til\mu<0$ is a chosen constant. At finite chemical potential $\mu<0$, superconductivity develops once the attraction is turned on. The weak-coupling regime manifests conventional BCS scaling, while the linear scaling is recovered at stronger interactions. Note that the critical interaction at $\mu=0$ is $V_c\sim1/m$. The true critical temperature $T_c$ of superconductivity exhibits similar behavior to $T_c^\txt{pair}$. While $T_c=T_c^\txt{pair}$ is manifest at weak coupling, the strong-coupling behavior of $T_c$ (\ref{eq:tc}) with linear scaling is determined from the competition between pair and phase fluctuations.}
\end{figure}

\subsubsection{Infinitesimal band curvature}

Our analysis has assumed a leading-order approximation (\ref{eq:ham0}) for the $\mbf k\cdot\mbf p$ Hamiltonian at band crossings $\pm P$. The middle band is perfectly flat under this assumption. In practice, however, the band structures usually involve higher-order corrections and according band curvatures. We thus introduce an infinitesimal quadratic perturbation $\d\mca H_{0\mbf k}=\sqrt6J_{0\mbf k}/2m$ with $L=S=2$ to the low-energy theory (\ref{eq:ham0}) \cite{lin18prb}. Such perturbation obeys the symmetry of low-energy theory, namely, spin-orbit coupled rotation and time-reversal symmetries. It is also compatible with the leading-order term $[\mca H_{0\mbf k},\d H_{0\mbf k}]=0$, so interband mixing does not occur. The flat band acquires a quadratic dispersion $\ve^0_{\mbf k}=-k^2/m$ under the perturbation [Fig.~\ref{fig:bandcurv}(a)]. The band curvature is assumed infinitesimal, $1/m\ll\L_\ve/\L_k^2$, so that the attractive regime still covers the whole (nearly) flat band. We place the chemical potential at a finite doping $-\L_k^2/m<\mu<0$, leading to a pair of spherical Fermi surfaces $\txt{FS}_\pm$ defined by Fermi momenta $|\mbf k_F|=(m\mu)^{1/2}$ around $\pm P$ (and infinitesimal shells from the hole linear bands). We further assume that the doping is far enough from the band crossing, so that the states $\ket{u^\pm_{\mbf k_F}}$ on the linear bands are far away from the attractive regime $|\ve^\pm_{\mbf k_F}-\mu|\gg\L_\ve$. This projects out the effect of multiband pairing on the Fermi surface \cite{agterberg17prl,venderbos18prx,brydon18prb}. The effective flat band theory (\ref{eq:bdg0}) under direct projection is thus eligible.

The energy scales of superconductivity can again be analyzed by solving the gap equation (\ref{eq:gapeq0}) \cite{lin18prb} and calculating the superfluid stiffness (\ref{eq:sfstf}). Due to the finite Fermi surface, superconductivity develops immediately as the interaction is turned on [Fig.~\ref{fig:bandcurv}(b)]. In the weak-coupling regime $V\ll1$, the finite density of states only yields the conventional BCS scaling. The critical temperature $T_c=T_c^\txt{pair}$ is determined by the pair fluctuations as in the conventional BCS theory. Nevertheless, the whole flat bands are involved in pairing at strong enough interactions. This enhances superconductivity dramatically and resumes the linear scalings (\ref{eq:linscal}) and (\ref{eq:tc}). The true critical temperature (\ref{eq:tc}) is determined from the competition between pair and phase fluctuations. Since the band curvature $1/m$ is infinitesimal, the required interaction for linear scaling is also infinitesimal.

\subsection{Flat band pairing channels}

We now examine each irreducible pairing channel and determine those with flat band superconductivity. Our analysis focuses on the channels with the first few angular momenta $J=0,1$, $L=0,1,2,3$, and $S=0,1,2$. These channels usually serve as the leading competitors in superconductivity. The orbital modes $L=0,1,2,3$ with respect to $\pm P$ are referred to as $s'$-, $p'$-, $d'$-, $f'$-wave pairings. Meanwhile, the effective orbital modes on flat bands are named $\bar s$-, $\bar p$-wave pairings in $J=0,1$ pairing channels. The results of our examination are summarized in Table \ref{tb:fbp}. It is worth noting that flat band pairing only occurs in spin singlet and quintet pairing channels. Spin triplet pairing does not support flat band pairing, since the according component in the spin representation vanishes: $\braket{u^0_{\mbf k}}{\vec S_1}{u^0_{\mbf k}}=0$.

The single-component $J=0$ pairing channels have been studied in previous analysis \cite{lin18prb}. Since $J=0$ only occurs when $L=S$, valley singlet pairing $\a=0$ is necessary under Fermi statistics. Flat band superconductivity can occur in $s'$-wave spin singlet and $d'$-wave spin quintet pairing channels $(L,S,J)=(0,0,0), (2,2,0)$. The according $\bar s$-wave gap functions are constant $\D_{\mbf k}\sim\D, (k_F/\L_k)^2\D$ on the Fermi surface, leading to fully gapped quasiparticle spectra. Valley singlet pairing imposes an opposite sign between the gap functions on $\txt{FS}_\pm$. This confirms the eligibility of pairing between nondegenerate flat bands under Fermi statistics.

Our main interest lies in the multicomponent $J=1$ pairing channels, which exhibit three-component order parameters $\vec\D$. By an exhaustive examination, we uncover flat band superconductivity in $p'$-wave spin singlet, $p'$-wave spin quintet, and $f'$-wave spin quintet pairing channels $(L,S,J)=(1,0,1), (1,2,1), (3,2,1)$ with valley triplet pairings $\a=3,\pm$. The gap functions in these pairing states take the $\bar p$-wave forms $\D_{\mbf k}\sim\vec\D\cdot\vec L_{1\mbf k}, (k_F/\L_k)^2\vec\D\cdot\vec L_{1\mbf k}$ on the Fermi surface. Note that each channel manifests a degenerate manifold of pairing states spanned by the three-component order parameter. Whichever type of pairing state is energetically favored determines the quasiparticle spectrum.

\begin{table}[t]
\centering
\betb{|c|c|c|c|c|c|}
\hline
$J$&$\a$&SOC&FBP&$(L,S,J)$&Gap on FS\\
\hline
\multirow{4}{*}{$0$}&\multirow{2}{*}{$0$}&\multirow{2}{*}{Even}&Yes&$(0,0,0)$, $(2,2,0)$& $\D_{\mbf k}\sim\D$\\\hhline{|~|~|~|---|}
&&&No&$(1,1,0)$&\cellcolor{gray}\\\hhline{|~|-----|}
&\multirow{2}{*}{$3$, $\pm$}&\multirow{2}{*}{Odd}&\cellcolor{gray}&\cellcolor{gray}&\cellcolor{gray}\\\hhline{|~|~|~|---|}
&&&\cellcolor{gray}&\cellcolor{gray}&\cellcolor{gray}\\
\hline
\multirow{4}{*}{$1$}&\multirow{2}{*}{$0$}&\multirow{2}{*}{Even}&Yes&\cellcolor{gray}&\cellcolor{gray}\\\hhline{|~|~|~|---|}
&&&No&$(1,1,1)$, $(2,2,1)$&\cellcolor{gray}\\\hhline{|~|-----|}
&\multirow{2}{*}{$3$, $\pm$}&\multirow{2}{*}{Odd}&Yes&$(1,0,1)$, $(1,2,1)$, $(3,2,1)$&$\D_{\mbf k}\sim\vec\D\cdot\vec L_{1\mbf k}$\\\hhline{|~|~|~|---|}
&&&No&$(0,1,1)$, $(2,1,1)$&\cellcolor{gray}\\
\hline
\entb
\caption{\label{tb:fbp} Exhaustive search for flat band pairing (FBP) in all irreducible pairing channels with $J=0,1$. Here SOC denotes the spin-orbit coupled pairing, which is combined with valley singlet (triplet) pairing $\a=0$ ($3,\pm$) when the exchange is even (odd). The $\bar s$-, $\bar p$-wave gap functions are manifest on the Fermi surface (FS) when flat band pairing occurs in $J=0,1$ pairing channels.}
\end{table}

\section{Chiral superconductivity with Majorana fermions}

We have found flat band superconductivity in the $J=1$ pairing channels with $\bar p$-wave pairings on the Fermi surface. These multicomponent pairing channels manifest degenerate manifolds which contain various pairing states. A natural question arises as which type of ground state is energetically favored when superconductivity develops. The according quasiparticle spectrum may host novel characteristics which are absent in single-component pairing states.

\subsection{Chiral ground states}

When superconductivity develops in the $J=1$ pairing channels, the potential ground states may fall into two distinct classes \cite{venderbos18prx}. The first class manifests chiral pairing orders, which carry finite axial angular momenta $M_J=\pm1$ and break time-reversal symmetry spontaneously. Meanwhile, the second class exhibits polar pairing orders, where $M_J=0$ and time-reversal symmetry is preserved. Whichever class is energetically favored depends on how the free energy achieves its minima in the manifold of $J=1$ pairing states.

The mean-field free energy can be derived from a coherent path integral calculation $f^0=|\mbf\D|^2/V-\Tr\ln(\mca G^0)^{-1}$ \cite{lin18prb}. Here $(\mca G^0)^{-1}_{\mbf kn}=i\o_n-\mca H^0_{\txt{BdG},\mbf k}$ is the inverse Gor'kov Green's function with fermionic Matsubara frequency $\o_n=(2n+1)\pi T$ and momentum $\mbf k$. For later convenience, we define the diagonal components as $G_{\pm,\mbf kn}=[i\o_n\mp(\ve^0_{\mbf k}-\mu)]^{-1}$. The trace in the free energy $f^0$ denotes a combination of matrix trace in Nambu space and frequency-momentum summation $(T/\mca V)\sum_{\mbf kn}$. Note that we have adopted the spatial representation of order parameter $\mbf\D=(\D_x,\D_y,\D_z)$, with the components obtained from $\D_{\pm1}=\mp(\D_x\mp i\D_y)/\sqrt2$ and $\D_0=\D_z$ under the relation $\vec\D\sim\vec J_{\mbf k}^\dag$. The $\bar p$-wave gap functions are expressed in a universal form $\D_{\mbf k}=c(\mbf k/\L_k)\cdot\mbf\D$ on the Fermi surface, where $c$ is a channel-dependent constant.

At mean-field level, the free energy can be expanded with infinitesimal order parameter near $T_c^\txt{pair}$ \cite{altlandbook}. An expansion up to quartic order gives the Ginzburg-Landau free energy
\beeq
\label{eq:freeeng}
f^0=r|\mbf\D|^2+u|\mbf\D|^4-\fr{u}{3}|\bar{\mbf\D}\vec{\mca I}_1\mbf\D|^2.
\eneq
In accordance with the onset of pairing below $T_c^\txt{pair}$, the prefactor of quadratic term $r=1/V+(c^2/3)\Tr[(k/\L_k)^2G_{+}G_{-}]$ turns negative, $r<0$, and triggers finite order parameter $|\mbf\D|\neq0$. The according energetically favored ground states are determined by the quartic terms. At quartic order, the isotropic term $|\mbf\D|^4$ exhibits a positive prefactor $u=(c^4/10)\Tr[(k/\L_k)^4G_+^2G_-^2]$ and stabilizes the free energy. Importantly, an additional anisotropy exists and favors finite subsidiary order $|\bar{\mbf\D}\vec{\mca I}_1\mbf\D|\neq0$ with representations $(\mca I_{a})_{bc}=-i\e_{abc}$ \cite{venderbos18prx}. The subsidiary order $\bar{\mbf\D}\vec{\mca I}_1\mbf\D$ represents the `magnetic dipole moment' of superconductivity, thereby capturing the spontaneous breakdown of time-reversal symmetry. With finite magnetic dipole moments favored, we conclude that the energetically favored ground states are the chiral $\bar p\pm i\bar p$ pairing states with $M_J=\pm1$. Note that the mean-field analysis of Ginzburg-Landau free energy has not involved phase fluctuations, which may suppress flat band superconductivity and reduce the critical temperature $T_c$ (\ref{eq:tc}). Nevertheless, the favoring of chiral pairing states is not altered, which always occurs when $\bar p$-wave superconductivity develops below $T_c$.

The favoring of chiral pairing states can be justified from the BdG quasiparticle gap structure. To minimize the free energy, the energetically favored ground states should maximize the ordering energy by maximizing the gap structure. As the potential ground states in the $J=1$ pairing channels, the chiral and polar pairing states manifest the gap functions $\D_{\mbf k}\sim\D[(k_x\pm ik_y)/\sqrt2\L_k]$ and $\D(k_z/\L_k)$, respectively. Note that the rotation axis has been set as $\hat z$ in this representation. For the low-energy theory (\ref{eq:ham0}) with full rotation symmetry, $\hat z$ may point in an arbitrary direction. However, the projection of practical interaction usually involves anisotropy along $\pm\mbf P$, thereby fixing the axis as $\hat z=\mbf P/|\mbf P|$. In a chiral pairing state, the gap function vanishes at the north and south poles $\mbf K_\pm=\pm k_F\hat z$ on the Fermi surface. This imposes a pair of nodal points at $\mbf K_\pm$ in the quasiparticle spectra. On the other hand, a polar pairing state exhibits a vanishing gap function at $k_z=0$, which leads to a nodal line along the equatorial. We immediately find that the chiral pairing states exhibit `larger' gap structure than the polar pairing states. This identifies the chiral pairing states as the energetically favored ground states, consistent with the mean-field analysis of Ginzburg-Landau free energy.

\subsection{Bulk and surface Majorana fermions}

\begin{figure}[b]
\centering
\includegraphics[scale = 1]{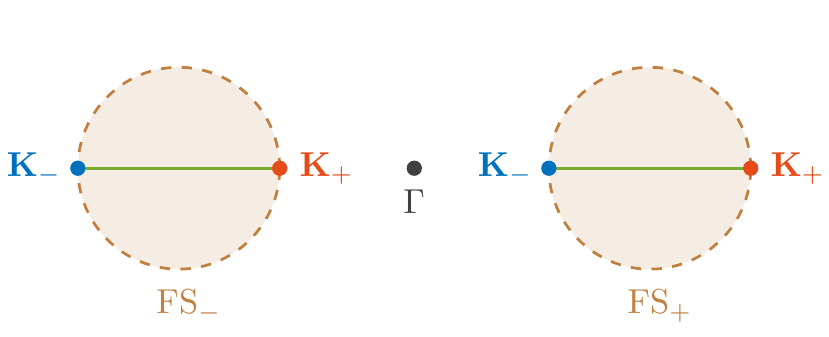}
\caption{\label{fig:cg} When the chiral $\bar p\pm i\bar p$ BCS state $\a=3$ develops, the Fermi surfaces $\txt{FS}_\pm$ around $\pm P$ are gapped out except at the bulk Majorana points $\mbf K_\pm$. Monopole charges $q_\pm=\pm1/2$ are carried by $\mbf K_\pm$'s, respectively. These bulk Majorana points bring about Majorana arcs in the surface Brillouin zone (green lines). The PDW state $\a=\pm$ only hosts bulk Majorana points on one Fermi surface $\txt{FS}_\pm$ and an according surface Majorana arc.}
\end{figure}

Chiral pairing states host novel features both in the bulk and on the surface. To uncover these features, we turn to the BdG Hamiltonian (\ref{eq:bdg0}) and study the quasiparticle spectrum. Consider the chiral $\bar p+i\bar p$ pairing state with the gap function
\beeq
\label{eq:chiralgapfun}
\D_{\mbf k}=-\fr{c\D}{\sqrt2\L_k}(k_x+ik_y).
\eneq
The quasiparticle spectrum is analogous to the one in the superfluid $^3$He-$A$ phase \cite{volovikbook}, where a pair of nodal points appears at $\mbf K_\pm$ (Fig.~\ref{fig:cg}). Note that the flat bands at three-band crossing points $\pm P$ do not host spin degeneracy as $^3$He does. While the spin-degenerate nodal points in the superfluid $^3$He-$A$ phase manifest low-energy Weyl quasiparticles, non-Weyl-type quasiparticles are expected at the nodal points herein.

To study the low-energy quasiparticles at the nodal points, we expand the BdG Hamiltonian (\ref{eq:bdg0}) in the vicinity of $\mbf K_\pm$ on $\txt{FS}_{\l_\pm}$. This yields a low-energy model $H^K=\sumv{q}\Psi_{\mbf q}^\dag\mca H_{\mbf q}\Psi_{\mbf q}$ of the four-component fermion $\Psi_{\mbf q}^\dag=(c_{\mbf q\l_++}^\dag,c_{\mbf q\l_--}^\dag,c_{-\mbf q\l_++},c_{-\mbf q\l_--})$ with $c_{\mbf q\l\pm}=c_{\mbf q+(\pm\mbf K)\l}$ \cite{kozii16sa}. The Hamiltonian
\beeq
\label{eq:majorana}
\mca H_{\mbf q}=-v_Fq_z\s^z-v_\D(q_x\tau^x-q_y\tau^y)\s^x
\eneq
exhibits linear dispersions $\pm E_{\mbf q}=\pm[v_\D^2(q_x^2+q_y^2)+v_F^2q_z^2]^{1/2}$ in the vicinity of $\mbf K_\pm$ on $\txt{FS}_{\l_\pm}$. Here $v_F=2k_F/m$ and $v_\D=c\D/\sqrt2\L_k$ are the effective velocities along and perpendicular to $\hat z$, respectively. The Pauli matrices are defined so that $\s^z=\pm1$ label $\mbf K_\pm$ on $\txt{FS}_{\l_\pm}$ and $\tau^z=\pm1$ denote particle-hole components. Remarkably, the four-component fermion is invariant under particle-hole transformation $\Psi_{\mbf q}^\dag=(\tau^x\Psi_{-\mbf q})^T$. This indicates the equivalence between any particle and its antiparticle in the low-energy model. The low-energy quasiparticles are thus identified as Majorana fermions \cite{kozii16sa}, which differ from the Weyl quasiparticles in the superfluid $^3$He-$A$ phase \cite{volovikbook}. Note that Hamiltonian (\ref{eq:majorana}) gives the real Majorana equation in relativistic quantum mechanics (with anisotropic velocity) \cite{elliott15rmp} when a `real' representation $\Psi_{\mbf q}^\dag=\Psi_{-\mbf q}$ is adopted. The Majorana feature is generic for spin-nondegenerate nodal points in 3D chiral superconductivity \cite{meng12prb,sau12prb,venderbos16prb,kozii16sa,venderbos18prx}. Unlike the Majorana bound states in one-dimensional (1D) and 2D chiral superconductivity \cite{read00prb}, the Majorana fermions herein are itinerant in the bulk of 3D chiral superconductivity.

Analogous to the Weyl points in Weyl semimetals \cite{armitage18rmp}, the bulk Majorana points carry nontrivial monopole charges. From the low-energy Hamiltonian (\ref{eq:majorana}), we obtain opposite monopole charges $q_\pm=\pm1/2$ at the Majorana points $\mbf K_\pm$ on $\txt{FS}_{\l_\pm}$. The numbers of Majorana points are different for different valley triplet pairings. In the BCS state $\a=3$, four Majorana points $\mbf K_\pm$ on $\txt{FS}_\pm$ are present (Fig.~\ref{fig:cg}). Valley triplet pairing imposes the same monopole charge $q_+=1/2$ at $\mbf K_+$'s on $\txt{FS}_\pm$. The other points $\mbf K_-$ carry an opposite monopole charge $q_-=-1/2$ to the one carried by $\mbf K_+$'s. The net vorticity on each Fermi surface $\txt{FS}_\pm$ is zero, as the flat band pairing does not exhibit nontrivial monopole structure. Such a configuration differs from the monopole superconductivity in inversion symmetric Weyl semimetals \cite{cho12prb,li18prl}. The latter manifests nonzero net vorticity on each Fermi surface due to nontrivial pairing monopole structure. On the other hand, the PDW states $\a=\pm$ manifest two Majorana points $\mbf K_\pm$ with opposite monopole charges on a single Fermi surface $\txt{FS}_\pm$.

The presence of bulk Majorana points generically leads to Majorana arcs in the surface Brillouin zone \cite{meng12prb,sau12prb,venderbos16prb,kozii16sa}. The configuration of these arcs depends on the surface of interest. Here we choose a surface parallel to the $xz$ plane, where $p_x$ and $p_z$ form the surface Brillouin zone. The surface zero mode at $p_z$ corresponds to the edge mode in the effective bulk 2D system $\mca H_{p_xp_y}^{\txt{2D},p_z}$ at $p_z$ \cite{sau12prb,armitage18rmp}. When $p_z$ lies between two Majorana points on the same Fermi surface $|p_z-(\pm P_z)|<k_F$, the 2D band encloses an odd number of Majorana points. Nontrivial Chern number $C=\pm1$ is manifested accordingly, leading to a topologically protected chiral edge mode at $p_x=0$. This edge mode is of Majorana type due to the bulk BdG structure. As $p_z$ goes into the rest region $|p_z-(\pm P_z)|>k_F$, the 2D band encloses pairs of Majorana points with opposite monopole charges. No topologically protected edge mode exists in this case. From these inspections, we conclude that Majorana bound states exist on the surface as open arcs. Each Majorana arc connects the projections of Majorana points $\mbf K_\pm$ from the same Fermi surface (Fig.~\ref{fig:cg}). The surface Majorana arcs may be probed experimentally by, for example, angle-resolved photoemission spectroscopy (ARPES), which is powerful in probing surface energy spectrum \cite{sau12prb}.

\section{Discussion}
\label{sec:concl}

We have studied the {\it spin-1 fermion} pairing states at a time-reversal pair of symmetry-protected three-band crossing points. Based on symmetry analysis, mean-field study, and superfluid stiffness calculation, we have exhaustively examined irreducible pairing channels with valley singlet, triplet, and spin-orbit coupled $J=0,1$ pairings. We have focused particularly on the channels with {\it flat band pairings}, where superconductivity can be dramatically enhanced. Such enhancement leads to a {\it linear} scaling of critical temperature in the interaction strength. While $J=0$ flat band pairing states exhibit full bulk gaps, we have uncovered $J=1$ {\it chiral $\bar p\pm i\bar p$ flat band superconductivity} with bulk topological nodal points. The spin-nondegenerate nodal points host {\it 3D itinerant Majorana fermions} as low-energy quasiparticles. Meanwhile, open {\it Majorana arcs} arise on the surface and connect the projections of bulk Majorana points.

Further investigations from this work are still open, as we briefly discuss below. Our work has analyzed the irreducible pairing channels on an equal footing, without addressing the issue of whichever channel is leading. We have also examined each channel independently, while interchannel intertwinement may occur in practice. The projection of practical interaction on low-energy theory may introduce explicit interchannel coupling, as well. Additionally, our analysis has adopted the full rotation symmetry of low-energy theory, while a reduction down to lattice group symmetry may alter the pairing states. Furthermore, the cutoff of the `flat band regime' has been left as undetermined in our analysis, where flat band enhancement may reduce and multiband pairing may arise. All of the above issues depend strongly on the details in the systems of interest. According analyses would provide useful information for the study of practical systems, which are left as future work. Meanwhile, our analysis has been conducted mostly at the mean-field level, without much exploration into the potential effects beyond this level. Examples of such effects include the variation of critical exponents due to strong phase fluctuations. This feature depends significantly on how the superconducting phase transition is affected by the strong phase fluctuations in three dimensions. While such effect may be analyzed with a 3D $XY$ model approximation, a well-recognized solution is still under investigation \cite{hazra19prx}. Another example is the retardation effect of mediating bosonic modes in the attractions, which has been studied in the Eliashberg theory \cite{marsiglio20ap}. In this scenario, the critical temperature may acquire a square-root scaling in the interaction strength $T_c\sim\sqrt{V}$ at strong coupling. Such behavior may be an interesting possibility for the flat band superconductivity at strong coupling, which goes beyond the framework of this work. Further investigations of these potential effects beyond the mean-field level may serve as interesting directions for future work. On the other hand, we have not addressed the fate of topologically nontrivial linear bands and according surface Fermi arcs. As flat band pairing develops at enhanced critical temperature, much weaker pairing also occurs on the infinitesimal Fermi surfaces of doped linear bands \cite{lin18prb}. Nontrivial topology may lead to remarkable features in the linear band pairing states, which also affects the behavior of normal Fermi arcs and connected additional (off-Fermi level) Weyl points \cite{wangye16prb}. Further investigation along this direction may be an interesting topic for future work. Finally, our analysis has focused on superconductivity without addressing the other instabilities. The study of the other potential flat band instabilities would be an interesting problem for future work.

Our work raises the interesting issue that dramatically enhanced chiral superconductivity can develop on 3D flat bands and host novel Majorana fermions. The information herein may be beneficial to the experiments on practical materials, and also to the theoretical study of novel superconductivity.

{\it Note added.} Recently, I learned about an independent study of superconductivity in systems with three-band crossings by Sim, Park, and Lee \cite{sim19ax}. While their work studies the valley triplet $s'$-wave spin triplet pairing channel at a broad range of doping, the analysis in this paper finds and focuses on other pairing channels with flat band pairing, which would support much stronger superconductivity in the vicinity of band crossings.

\begin{acknowledgments}
I acknowledge Daniel Agterberg, Wei-Han Hsiao, Sebastiano Peotta, J\"orn Venderbos, and Yuxuan Wang for valuable feedback and suggestions on the manuscript. I especially thank Rahul Nandkishore for stimulating discussions and encouragement on this work, as well as prior collaboration on a related work. This research was sponsored by the Army Research Office and was accomplished under Grant No. W911NF-17-1-0482. The views and conclusions contained in this document are those of the authors and should not be interpreted as representing the official policies, either expressed or implied, of the Army Research Office or the U.S. Government. The U.S. Government is authorized to reproduce and distribute reprints for Government purposes notwithstanding any copyright annotation thereon.
\end{acknowledgments}

\bibliography{Reference}

\begin{thebibliography}{58}%
\makeatletter
\providecommand \@ifxundefined [1]{%
 \@ifx{#1\undefined}
}%
\providecommand \@ifnum [1]{%
 \ifnum #1\expandafter \@firstoftwo
 \else \expandafter \@secondoftwo
 \fi
}%
\providecommand \@ifx [1]{%
 \ifx #1\expandafter \@firstoftwo
 \else \expandafter \@secondoftwo
 \fi
}%
\providecommand \natexlab [1]{#1}%
\providecommand \enquote  [1]{``#1''}%
\providecommand \bibnamefont  [1]{#1}%
\providecommand \bibfnamefont [1]{#1}%
\providecommand \citenamefont [1]{#1}%
\providecommand \href@noop [0]{\@secondoftwo}%
\providecommand \href [0]{\begingroup \@sanitize@url \@href}%
\providecommand \@href[1]{\@@startlink{#1}\@@href}%
\providecommand \@@href[1]{\endgroup#1\@@endlink}%
\providecommand \@sanitize@url [0]{\catcode `\\12\catcode `\$12\catcode
  `\&12\catcode `\#12\catcode `\^12\catcode `\_12\catcode `\%12\relax}%
\providecommand \@@startlink[1]{}%
\providecommand \@@endlink[0]{}%
\providecommand \url  [0]{\begingroup\@sanitize@url \@url }%
\providecommand \@url [1]{\endgroup\@href {#1}{\urlprefix }}%
\providecommand \urlprefix  [0]{URL }%
\providecommand \Eprint [0]{\href }%
\providecommand \doibase [0]{http://dx.doi.org/}%
\providecommand \selectlanguage [0]{\@gobble}%
\providecommand \bibinfo  [0]{\@secondoftwo}%
\providecommand \bibfield  [0]{\@secondoftwo}%
\providecommand \translation [1]{[#1]}%
\providecommand \BibitemOpen [0]{}%
\providecommand \bibitemStop [0]{}%
\providecommand \bibitemNoStop [0]{.\EOS\space}%
\providecommand \EOS [0]{\spacefactor3000\relax}%
\providecommand \BibitemShut  [1]{\csname bibitem#1\endcsname}%
\let\auto@bib@innerbib\@empty
\bibitem [{\citenamefont {Kallin}\ and\ \citenamefont
  {Berlinsky}(2016)}]{kallin16rpp}%
  \BibitemOpen
  \bibfield  {author} {\bibinfo {author} {\bibfnamefont {C.}~\bibnamefont
  {Kallin}}\ and\ \bibinfo {author} {\bibfnamefont {J.}~\bibnamefont
  {Berlinsky}},\ }\bibfield  {title} {\enquote {\bibinfo {title} {Chiral
  superconductors},}\ }\href {\doibase 10.1088/0034-4885/79/5/054502}
  {\bibfield  {journal} {\bibinfo  {journal} {Rep. Prog. Phys.}\ }\textbf
  {\bibinfo {volume} {79}},\ \bibinfo {pages} {054502} (\bibinfo {year}
  {2016})}\BibitemShut {NoStop}%
\bibitem [{\citenamefont {Sato}\ and\ \citenamefont {Ando}(2017)}]{sato17rpp}%
  \BibitemOpen
  \bibfield  {author} {\bibinfo {author} {\bibfnamefont {M.}~\bibnamefont
  {Sato}}\ and\ \bibinfo {author} {\bibfnamefont {Y.}~\bibnamefont {Ando}},\
  }\bibfield  {title} {\enquote {\bibinfo {title} {Topological superconductors:
  a review},}\ }\href {\doibase 10.1088/1361-6633/aa6ac7} {\bibfield  {journal}
  {\bibinfo  {journal} {Rep. Prog. Phys.}\ }\textbf {\bibinfo {volume} {80}},\
  \bibinfo {pages} {076501} (\bibinfo {year} {2017})}\BibitemShut {NoStop}%
\bibitem [{\citenamefont {Read}\ and\ \citenamefont {Green}(2000)}]{read00prb}%
  \BibitemOpen
  \bibfield  {author} {\bibinfo {author} {\bibfnamefont {N.}~\bibnamefont
  {Read}}\ and\ \bibinfo {author} {\bibfnamefont {D.}~\bibnamefont {Green}},\
  }\bibfield  {title} {\enquote {\bibinfo {title} {Paired states of fermions in
  two dimensions with breaking of parity and time-reversal symmetries and the
  fractional quantum {Hall} effect},}\ }\href {\doibase
  10.1103/PhysRevB.61.10267} {\bibfield  {journal} {\bibinfo  {journal} {Phys.
  Rev. B}\ }\textbf {\bibinfo {volume} {61}},\ \bibinfo {pages} {10267}
  (\bibinfo {year} {2000})}\BibitemShut {NoStop}%
\bibitem [{\citenamefont {Mackenzie}\ and\ \citenamefont
  {Maeno}(2003)}]{mackenzie03rmp}%
  \BibitemOpen
  \bibfield  {author} {\bibinfo {author} {\bibfnamefont {A.~P.}\ \bibnamefont
  {Mackenzie}}\ and\ \bibinfo {author} {\bibfnamefont {Y.}~\bibnamefont
  {Maeno}},\ }\bibfield  {title} {\enquote {\bibinfo {title} {The
  superconductivity of {${\mathrm{Sr}}_{2}{\mathrm{RuO}}_{4}$} and the physics
  of spin-triplet pairing},}\ }\href {\doibase 10.1103/RevModPhys.75.657}
  {\bibfield  {journal} {\bibinfo  {journal} {Rev. Mod. Phys.}\ }\textbf
  {\bibinfo {volume} {75}},\ \bibinfo {pages} {657} (\bibinfo {year}
  {2003})}\BibitemShut {NoStop}%
\bibitem [{\citenamefont {Fischer}\ \emph {et~al.}(2014)\citenamefont
  {Fischer}, \citenamefont {Neupert}, \citenamefont {Platt}, \citenamefont
  {Schnyder}, \citenamefont {Hanke}, \citenamefont {Goryo}, \citenamefont
  {Thomale},\ and\ \citenamefont {Sigrist}}]{fischer14prb}%
  \BibitemOpen
  \bibfield  {author} {\bibinfo {author} {\bibfnamefont {M.~H.}\ \bibnamefont
  {Fischer}}, \bibinfo {author} {\bibfnamefont {T.}~\bibnamefont {Neupert}},
  \bibinfo {author} {\bibfnamefont {C.}~\bibnamefont {Platt}}, \bibinfo
  {author} {\bibfnamefont {A.~P.}\ \bibnamefont {Schnyder}}, \bibinfo {author}
  {\bibfnamefont {W.}~\bibnamefont {Hanke}}, \bibinfo {author} {\bibfnamefont
  {J.}~\bibnamefont {Goryo}}, \bibinfo {author} {\bibfnamefont
  {R.}~\bibnamefont {Thomale}}, \ and\ \bibinfo {author} {\bibfnamefont
  {M.}~\bibnamefont {Sigrist}},\ }\bibfield  {title} {\enquote {\bibinfo
  {title} {Chiral $d$-wave superconductivity in {SrPtAs}},}\ }\href {\doibase
  10.1103/PhysRevB.89.020509} {\bibfield  {journal} {\bibinfo  {journal} {Phys.
  Rev. B}\ }\textbf {\bibinfo {volume} {89}},\ \bibinfo {pages} {020509}
  (\bibinfo {year} {2014})}\BibitemShut {NoStop}%
\bibitem [{\citenamefont {Nandkishore}\ \emph {et~al.}(2012)\citenamefont
  {Nandkishore}, \citenamefont {Levitov},\ and\ \citenamefont
  {Chubukov}}]{nandkishore12np}%
  \BibitemOpen
  \bibfield  {author} {\bibinfo {author} {\bibfnamefont {R.}~\bibnamefont
  {Nandkishore}}, \bibinfo {author} {\bibfnamefont {L.~S.}\ \bibnamefont
  {Levitov}}, \ and\ \bibinfo {author} {\bibfnamefont {A.~V.}\ \bibnamefont
  {Chubukov}},\ }\bibfield  {title} {\enquote {\bibinfo {title} {Chiral
  superconductivity from repulsive interactions in doped graphene},}\ }\href
  {\doibase 10.1038/nphys2208} {\bibfield  {journal} {\bibinfo  {journal} {Nat.
  Phys.}\ }\textbf {\bibinfo {volume} {8}},\ \bibinfo {pages} {158} (\bibinfo
  {year} {2012})}\BibitemShut {NoStop}%
\bibitem [{\citenamefont {Lin}\ and\ \citenamefont
  {Nandkishore}(2019)}]{lin19prb}%
  \BibitemOpen
  \bibfield  {author} {\bibinfo {author} {\bibfnamefont {Y.-P.}\ \bibnamefont
  {Lin}}\ and\ \bibinfo {author} {\bibfnamefont {R.~M.}\ \bibnamefont
  {Nandkishore}},\ }\bibfield  {title} {\enquote {\bibinfo {title} {Chiral
  twist on the high-${T}_{c}$ phase diagram in moir\'e heterostructures},}\
  }\href {\doibase 10.1103/PhysRevB.100.085136} {\bibfield  {journal} {\bibinfo
   {journal} {Phys. Rev. B}\ }\textbf {\bibinfo {volume} {100}},\ \bibinfo
  {pages} {085136} (\bibinfo {year} {2019})}\BibitemShut {NoStop}%
\bibitem [{\citenamefont {Son}(2018)}]{son18anncmp}%
  \BibitemOpen
  \bibfield  {author} {\bibinfo {author} {\bibfnamefont {D.~T.}\ \bibnamefont
  {Son}},\ }\bibfield  {title} {\enquote {\bibinfo {title} {The {Dirac}
  composite fermion of the fractional quantum {Hall} effect},}\ }\href
  {\doibase 10.1146/annurev-conmatphys-033117-054227} {\bibfield  {journal}
  {\bibinfo  {journal} {Annu. Rev. Condens. Matter Phys.}\ }\textbf {\bibinfo
  {volume} {9}},\ \bibinfo {pages} {397} (\bibinfo {year} {2018})}\BibitemShut
  {NoStop}%
\bibitem [{\citenamefont {Sigrist}\ and\ \citenamefont
  {Ueda}(1991)}]{sigrist91rmp}%
  \BibitemOpen
  \bibfield  {author} {\bibinfo {author} {\bibfnamefont {M.}~\bibnamefont
  {Sigrist}}\ and\ \bibinfo {author} {\bibfnamefont {K.}~\bibnamefont {Ueda}},\
  }\bibfield  {title} {\enquote {\bibinfo {title} {Phenomenological theory of
  unconventional superconductivity},}\ }\href {\doibase
  10.1103/RevModPhys.63.239} {\bibfield  {journal} {\bibinfo  {journal} {Rev.
  Mod. Phys.}\ }\textbf {\bibinfo {volume} {63}},\ \bibinfo {pages} {239}
  (\bibinfo {year} {1991})}\BibitemShut {NoStop}%
\bibitem [{\citenamefont {Nandkishore}(2016)}]{nandkishore16prb}%
  \BibitemOpen
  \bibfield  {author} {\bibinfo {author} {\bibfnamefont {R.}~\bibnamefont
  {Nandkishore}},\ }\bibfield  {title} {\enquote {\bibinfo {title} {Weyl and
  {Dirac} loop superconductors},}\ }\href {\doibase 10.1103/PhysRevB.93.020506}
  {\bibfield  {journal} {\bibinfo  {journal} {Phys. Rev. B}\ }\textbf {\bibinfo
  {volume} {93}},\ \bibinfo {pages} {020506} (\bibinfo {year}
  {2016})}\BibitemShut {NoStop}%
\bibitem [{\citenamefont {Agterberg}\ \emph {et~al.}(2017)\citenamefont
  {Agterberg}, \citenamefont {Brydon},\ and\ \citenamefont
  {Timm}}]{agterberg17prl}%
  \BibitemOpen
  \bibfield  {author} {\bibinfo {author} {\bibfnamefont {D.~F.}\ \bibnamefont
  {Agterberg}}, \bibinfo {author} {\bibfnamefont {P.~M.~R.}\ \bibnamefont
  {Brydon}}, \ and\ \bibinfo {author} {\bibfnamefont {C.}~\bibnamefont
  {Timm}},\ }\bibfield  {title} {\enquote {\bibinfo {title} {Bogoliubov {Fermi}
  surfaces in superconductors with broken time-reversal symmetry},}\ }\href
  {\doibase 10.1103/PhysRevLett.118.127001} {\bibfield  {journal} {\bibinfo
  {journal} {Phys. Rev. Lett.}\ }\textbf {\bibinfo {volume} {118}},\ \bibinfo
  {pages} {127001} (\bibinfo {year} {2017})}\BibitemShut {NoStop}%
\bibitem [{\citenamefont {Brydon}\ \emph {et~al.}(2018)\citenamefont {Brydon},
  \citenamefont {Agterberg}, \citenamefont {Menke},\ and\ \citenamefont
  {Timm}}]{brydon18prb}%
  \BibitemOpen
  \bibfield  {author} {\bibinfo {author} {\bibfnamefont {P.~M.~R.}\
  \bibnamefont {Brydon}}, \bibinfo {author} {\bibfnamefont {D.~F.}\
  \bibnamefont {Agterberg}}, \bibinfo {author} {\bibfnamefont {H.}~\bibnamefont
  {Menke}}, \ and\ \bibinfo {author} {\bibfnamefont {C.}~\bibnamefont {Timm}},\
  }\bibfield  {title} {\enquote {\bibinfo {title} {Bogoliubov {Fermi} surfaces:
  General theory, magnetic order, and topology},}\ }\href {\doibase
  10.1103/PhysRevB.98.224509} {\bibfield  {journal} {\bibinfo  {journal} {Phys.
  Rev. B}\ }\textbf {\bibinfo {volume} {98}},\ \bibinfo {pages} {224509}
  (\bibinfo {year} {2018})}\BibitemShut {NoStop}%
\bibitem [{\citenamefont {Venderbos}\ \emph {et~al.}(2018)\citenamefont
  {Venderbos}, \citenamefont {Savary}, \citenamefont {Ruhman}, \citenamefont
  {Lee},\ and\ \citenamefont {Fu}}]{venderbos18prx}%
  \BibitemOpen
  \bibfield  {author} {\bibinfo {author} {\bibfnamefont {J.~W.~F.}\
  \bibnamefont {Venderbos}}, \bibinfo {author} {\bibfnamefont {L.}~\bibnamefont
  {Savary}}, \bibinfo {author} {\bibfnamefont {J.}~\bibnamefont {Ruhman}},
  \bibinfo {author} {\bibfnamefont {P.~A.}\ \bibnamefont {Lee}}, \ and\
  \bibinfo {author} {\bibfnamefont {L.}~\bibnamefont {Fu}},\ }\bibfield
  {title} {\enquote {\bibinfo {title} {Pairing states of spin-$\frac{3}{2}$
  fermions: Symmetry-enforced topological gap functions},}\ }\href {\doibase
  10.1103/PhysRevX.8.011029} {\bibfield  {journal} {\bibinfo  {journal} {Phys.
  Rev. X}\ }\textbf {\bibinfo {volume} {8}},\ \bibinfo {pages} {011029}
  (\bibinfo {year} {2018})}\BibitemShut {NoStop}%
\bibitem [{\citenamefont {Volovik}(2003)}]{volovikbook}%
  \BibitemOpen
  \bibfield  {author} {\bibinfo {author} {\bibfnamefont {G.~E.}\ \bibnamefont
  {Volovik}},\ }\href@noop {} {\emph {\bibinfo {title} {The Universe in a
  Helium Droplet}}}\ (\bibinfo  {publisher} {Oxford University Press, New
  York},\ \bibinfo {year} {2003})\BibitemShut {NoStop}%
\bibitem [{\citenamefont {Joynt}\ and\ \citenamefont
  {Taillefer}(2002)}]{joynt02rmp}%
  \BibitemOpen
  \bibfield  {author} {\bibinfo {author} {\bibfnamefont {R.}~\bibnamefont
  {Joynt}}\ and\ \bibinfo {author} {\bibfnamefont {L.}~\bibnamefont
  {Taillefer}},\ }\bibfield  {title} {\enquote {\bibinfo {title} {The
  superconducting phases of {${\mathrm{UPt}}_{3}$}},}\ }\href {\doibase
  10.1103/RevModPhys.74.235} {\bibfield  {journal} {\bibinfo  {journal} {Rev.
  Mod. Phys.}\ }\textbf {\bibinfo {volume} {74}},\ \bibinfo {pages} {235}
  (\bibinfo {year} {2002})}\BibitemShut {NoStop}%
\bibitem [{\citenamefont {Brydon}\ \emph {et~al.}(2016)\citenamefont {Brydon},
  \citenamefont {Wang}, \citenamefont {Weinert},\ and\ \citenamefont
  {Agterberg}}]{brydon16prl}%
  \BibitemOpen
  \bibfield  {author} {\bibinfo {author} {\bibfnamefont {P.~M.~R.}\
  \bibnamefont {Brydon}}, \bibinfo {author} {\bibfnamefont {L.}~\bibnamefont
  {Wang}}, \bibinfo {author} {\bibfnamefont {M.}~\bibnamefont {Weinert}}, \
  and\ \bibinfo {author} {\bibfnamefont {D.~F.}\ \bibnamefont {Agterberg}},\
  }\bibfield  {title} {\enquote {\bibinfo {title} {Pairing of $j=3/2$ fermions
  in half-{Heusler} superconductors},}\ }\href {\doibase
  10.1103/PhysRevLett.116.177001} {\bibfield  {journal} {\bibinfo  {journal}
  {Phys. Rev. Lett.}\ }\textbf {\bibinfo {volume} {116}},\ \bibinfo {pages}
  {177001} (\bibinfo {year} {2016})}\BibitemShut {NoStop}%
\bibitem [{\citenamefont {Boettcher}\ and\ \citenamefont
  {Herbut}(2018)}]{boettcher18prl}%
  \BibitemOpen
  \bibfield  {author} {\bibinfo {author} {\bibfnamefont {I.}~\bibnamefont
  {Boettcher}}\ and\ \bibinfo {author} {\bibfnamefont {I.~F.}\ \bibnamefont
  {Herbut}},\ }\bibfield  {title} {\enquote {\bibinfo {title} {Unconventional
  superconductivity in {Luttinger} semimetals: Theory of complex tensor order
  and the emergence of the uniaxial nematic state},}\ }\href {\doibase
  10.1103/PhysRevLett.120.057002} {\bibfield  {journal} {\bibinfo  {journal}
  {Phys. Rev. Lett.}\ }\textbf {\bibinfo {volume} {120}},\ \bibinfo {pages}
  {057002} (\bibinfo {year} {2018})}\BibitemShut {NoStop}%
\bibitem [{\citenamefont {Roy}\ \emph {et~al.}(2019)\citenamefont {Roy},
  \citenamefont {Ghorashi}, \citenamefont {Foster},\ and\ \citenamefont
  {Nevidomskyy}}]{roy99prb}%
  \BibitemOpen
  \bibfield  {author} {\bibinfo {author} {\bibfnamefont {B.}~\bibnamefont
  {Roy}}, \bibinfo {author} {\bibfnamefont {S.~A.~A.}\ \bibnamefont
  {Ghorashi}}, \bibinfo {author} {\bibfnamefont {M.~S.}\ \bibnamefont
  {Foster}}, \ and\ \bibinfo {author} {\bibfnamefont {A.~H.}\ \bibnamefont
  {Nevidomskyy}},\ }\bibfield  {title} {\enquote {\bibinfo {title} {Topological
  superconductivity of spin-$3/2$ carriers in a three-dimensional doped
  {Luttinger} semimetal},}\ }\href {\doibase 10.1103/PhysRevB.99.054505}
  {\bibfield  {journal} {\bibinfo  {journal} {Phys. Rev. B}\ }\textbf {\bibinfo
  {volume} {99}},\ \bibinfo {pages} {054505} (\bibinfo {year}
  {2019})}\BibitemShut {NoStop}%
\bibitem [{\citenamefont {Lee}\ \emph {et~al.}(2006)\citenamefont {Lee},
  \citenamefont {Nagaosa},\ and\ \citenamefont {Wen}}]{lee06rmp}%
  \BibitemOpen
  \bibfield  {author} {\bibinfo {author} {\bibfnamefont {P.~A.}\ \bibnamefont
  {Lee}}, \bibinfo {author} {\bibfnamefont {N.}~\bibnamefont {Nagaosa}}, \ and\
  \bibinfo {author} {\bibfnamefont {X.-G.}\ \bibnamefont {Wen}},\ }\bibfield
  {title} {\enquote {\bibinfo {title} {Doping a {Mott} insulator: Physics of
  high-temperature superconductivity},}\ }\href {\doibase
  10.1103/RevModPhys.78.17} {\bibfield  {journal} {\bibinfo  {journal} {Rev.
  Mod. Phys.}\ }\textbf {\bibinfo {volume} {78}},\ \bibinfo {pages} {17}
  (\bibinfo {year} {2006})}\BibitemShut {NoStop}%
\bibitem [{\citenamefont {Miyahara}\ \emph {et~al.}(2007)\citenamefont
  {Miyahara}, \citenamefont {Kusuta},\ and\ \citenamefont
  {Furukawa}}]{miyahara07pc}%
  \BibitemOpen
  \bibfield  {author} {\bibinfo {author} {\bibfnamefont {S.}~\bibnamefont
  {Miyahara}}, \bibinfo {author} {\bibfnamefont {S.}~\bibnamefont {Kusuta}}, \
  and\ \bibinfo {author} {\bibfnamefont {N.}~\bibnamefont {Furukawa}},\
  }\bibfield  {title} {\enquote {\bibinfo {title} {{BCS} theory on a flat band
  lattice},}\ }\href {\doibase https://doi.org/10.1016/j.physc.2007.03.393}
  {\bibfield  {journal} {\bibinfo  {journal} {Physica C: Superconductivity}\
  }\textbf {\bibinfo {volume} {460}},\ \bibinfo {pages} {1145 } (\bibinfo
  {year} {2007})}\BibitemShut {NoStop}%
\bibitem [{\citenamefont {Kopnin}\ \emph {et~al.}(2011)\citenamefont {Kopnin},
  \citenamefont {Heikkil\"a},\ and\ \citenamefont {Volovik}}]{kopnin11prb}%
  \BibitemOpen
  \bibfield  {author} {\bibinfo {author} {\bibfnamefont {N.~B.}\ \bibnamefont
  {Kopnin}}, \bibinfo {author} {\bibfnamefont {T.~T.}\ \bibnamefont
  {Heikkil\"a}}, \ and\ \bibinfo {author} {\bibfnamefont {G.~E.}\ \bibnamefont
  {Volovik}},\ }\bibfield  {title} {\enquote {\bibinfo {title}
  {High-temperature surface superconductivity in topological flat-band
  systems},}\ }\href {\doibase 10.1103/PhysRevB.83.220503} {\bibfield
  {journal} {\bibinfo  {journal} {Phys. Rev. B}\ }\textbf {\bibinfo {volume}
  {83}},\ \bibinfo {pages} {220503} (\bibinfo {year} {2011})}\BibitemShut
  {NoStop}%
\bibitem [{\citenamefont {Heikkil{\"a}}\ \emph {et~al.}(2011)\citenamefont
  {Heikkil{\"a}}, \citenamefont {Kopnin},\ and\ \citenamefont
  {Volovik}}]{heikkila11jetp}%
  \BibitemOpen
  \bibfield  {author} {\bibinfo {author} {\bibfnamefont {T.~T.}\ \bibnamefont
  {Heikkil{\"a}}}, \bibinfo {author} {\bibfnamefont {N.~B.}\ \bibnamefont
  {Kopnin}}, \ and\ \bibinfo {author} {\bibfnamefont {G.~E.}\ \bibnamefont
  {Volovik}},\ }\bibfield  {title} {\enquote {\bibinfo {title} {Flat bands in
  topological media},}\ }\href {\doibase 10.1134/S0021364011150045} {\bibfield
  {journal} {\bibinfo  {journal} {JETP Lett.}\ }\textbf {\bibinfo {volume}
  {94}},\ \bibinfo {pages} {233} (\bibinfo {year} {2011})}\BibitemShut
  {NoStop}%
\bibitem [{\citenamefont {Uchoa}\ and\ \citenamefont
  {Barlas}(2013)}]{uchoa13prl}%
  \BibitemOpen
  \bibfield  {author} {\bibinfo {author} {\bibfnamefont {B.}~\bibnamefont
  {Uchoa}}\ and\ \bibinfo {author} {\bibfnamefont {Y.}~\bibnamefont {Barlas}},\
  }\bibfield  {title} {\enquote {\bibinfo {title} {Superconducting states in
  pseudo-{Landau}-levels of strained graphene},}\ }\href {\doibase
  10.1103/PhysRevLett.111.046604} {\bibfield  {journal} {\bibinfo  {journal}
  {Phys. Rev. Lett.}\ }\textbf {\bibinfo {volume} {111}},\ \bibinfo {pages}
  {046604} (\bibinfo {year} {2013})}\BibitemShut {NoStop}%
\bibitem [{\citenamefont {Lin}\ and\ \citenamefont
  {Nandkishore}(2018)}]{lin18prb}%
  \BibitemOpen
  \bibfield  {author} {\bibinfo {author} {\bibfnamefont {Y.-P.}\ \bibnamefont
  {Lin}}\ and\ \bibinfo {author} {\bibfnamefont {R.~M.}\ \bibnamefont
  {Nandkishore}},\ }\bibfield  {title} {\enquote {\bibinfo {title} {Exotic
  superconductivity with enhanced energy scales in materials with three band
  crossings},}\ }\href {\doibase 10.1103/PhysRevB.97.134521} {\bibfield
  {journal} {\bibinfo  {journal} {Phys. Rev. B}\ }\textbf {\bibinfo {volume}
  {97}},\ \bibinfo {pages} {134521} (\bibinfo {year} {2018})}\BibitemShut
  {NoStop}%
\bibitem [{\citenamefont {Peotta}\ and\ \citenamefont
  {T\"orm\"a}(2015)}]{peotta15nc}%
  \BibitemOpen
  \bibfield  {author} {\bibinfo {author} {\bibfnamefont {S.}~\bibnamefont
  {Peotta}}\ and\ \bibinfo {author} {\bibfnamefont {P.}~\bibnamefont
  {T\"orm\"a}},\ }\bibfield  {title} {\enquote {\bibinfo {title} {Superfluidity
  in topologically nontrivial flat bands},}\ }\href {\doibase
  10.1038/ncomms9944} {\bibfield  {journal} {\bibinfo  {journal} {Nat.
  Commun.}\ }\textbf {\bibinfo {volume} {6}},\ \bibinfo {pages} {8944}
  (\bibinfo {year} {2015})}\BibitemShut {NoStop}%
\bibitem [{\citenamefont {Julku}\ \emph {et~al.}(2016)\citenamefont {Julku},
  \citenamefont {Peotta}, \citenamefont {Vanhala}, \citenamefont {Kim},\ and\
  \citenamefont {T\"orm\"a}}]{julku16prl}%
  \BibitemOpen
  \bibfield  {author} {\bibinfo {author} {\bibfnamefont {A.}~\bibnamefont
  {Julku}}, \bibinfo {author} {\bibfnamefont {S.}~\bibnamefont {Peotta}},
  \bibinfo {author} {\bibfnamefont {T.~I.}\ \bibnamefont {Vanhala}}, \bibinfo
  {author} {\bibfnamefont {D.-H.}\ \bibnamefont {Kim}}, \ and\ \bibinfo
  {author} {\bibfnamefont {P.}~\bibnamefont {T\"orm\"a}},\ }\bibfield  {title}
  {\enquote {\bibinfo {title} {Geometric origin of superfluidity in the
  {Lieb}-lattice flat band},}\ }\href {\doibase 10.1103/PhysRevLett.117.045303}
  {\bibfield  {journal} {\bibinfo  {journal} {Phys. Rev. Lett.}\ }\textbf
  {\bibinfo {volume} {117}},\ \bibinfo {pages} {045303} (\bibinfo {year}
  {2016})}\BibitemShut {NoStop}%
\bibitem [{\citenamefont {Liang}\ \emph {et~al.}(2017)\citenamefont {Liang},
  \citenamefont {Vanhala}, \citenamefont {Peotta}, \citenamefont {Siro},
  \citenamefont {Harju},\ and\ \citenamefont {T\"orm\"a}}]{liang17prb}%
  \BibitemOpen
  \bibfield  {author} {\bibinfo {author} {\bibfnamefont {L.}~\bibnamefont
  {Liang}}, \bibinfo {author} {\bibfnamefont {T.~I.}\ \bibnamefont {Vanhala}},
  \bibinfo {author} {\bibfnamefont {S.}~\bibnamefont {Peotta}}, \bibinfo
  {author} {\bibfnamefont {T.}~\bibnamefont {Siro}}, \bibinfo {author}
  {\bibfnamefont {A.}~\bibnamefont {Harju}}, \ and\ \bibinfo {author}
  {\bibfnamefont {P.}~\bibnamefont {T\"orm\"a}},\ }\bibfield  {title} {\enquote
  {\bibinfo {title} {Band geometry, berry curvature, and superfluid weight},}\
  }\href {\doibase 10.1103/PhysRevB.95.024515} {\bibfield  {journal} {\bibinfo
  {journal} {Phys. Rev. B}\ }\textbf {\bibinfo {volume} {95}},\ \bibinfo
  {pages} {024515} (\bibinfo {year} {2017})}\BibitemShut {NoStop}%
\bibitem [{\citenamefont {Xie}\ \emph {et~al.}(2020)\citenamefont {Xie},
  \citenamefont {Song}, \citenamefont {Lian},\ and\ \citenamefont
  {Bernevig}}]{xie20prl}%
  \BibitemOpen
  \bibfield  {author} {\bibinfo {author} {\bibfnamefont {F.}~\bibnamefont
  {Xie}}, \bibinfo {author} {\bibfnamefont {Z.}~\bibnamefont {Song}}, \bibinfo
  {author} {\bibfnamefont {B.}~\bibnamefont {Lian}}, \ and\ \bibinfo {author}
  {\bibfnamefont {B.~A.}\ \bibnamefont {Bernevig}},\ }\bibfield  {title}
  {\enquote {\bibinfo {title} {Topology-bounded superfluid weight in twisted
  bilayer graphene},}\ }\href {\doibase 10.1103/PhysRevLett.124.167002}
  {\bibfield  {journal} {\bibinfo  {journal} {Phys. Rev. Lett.}\ }\textbf
  {\bibinfo {volume} {124}},\ \bibinfo {pages} {167002} (\bibinfo {year}
  {2020})}\BibitemShut {NoStop}%
\bibitem [{\citenamefont {Cao}\ \emph {et~al.}(2018)\citenamefont {Cao},
  \citenamefont {Fatemi}, \citenamefont {Fang}, \citenamefont {Watanabe},
  \citenamefont {Taniguchi}, \citenamefont {Kaxiras},\ and\ \citenamefont
  {Jarillo-Herrero}}]{cao18n}%
  \BibitemOpen
  \bibfield  {author} {\bibinfo {author} {\bibfnamefont {Y.}~\bibnamefont
  {Cao}}, \bibinfo {author} {\bibfnamefont {V.}~\bibnamefont {Fatemi}},
  \bibinfo {author} {\bibfnamefont {S.}~\bibnamefont {Fang}}, \bibinfo {author}
  {\bibfnamefont {K.}~\bibnamefont {Watanabe}}, \bibinfo {author}
  {\bibfnamefont {T.}~\bibnamefont {Taniguchi}}, \bibinfo {author}
  {\bibfnamefont {E.}~\bibnamefont {Kaxiras}}, \ and\ \bibinfo {author}
  {\bibfnamefont {P.}~\bibnamefont {Jarillo-Herrero}},\ }\bibfield  {title}
  {\enquote {\bibinfo {title} {Unconventional superconductivity in magic-angle
  graphene superlattices},}\ }\href {\doibase 10.1038/nature26160} {\bibfield
  {journal} {\bibinfo  {journal} {Nature}\ }\textbf {\bibinfo {volume} {556}},\
  \bibinfo {pages} {43} (\bibinfo {year} {2018})}\BibitemShut {NoStop}%
\bibitem [{\citenamefont {Chen}\ \emph {et~al.}(2019)\citenamefont {Chen},
  \citenamefont {Sharpe}, \citenamefont {Gallagher}, \citenamefont {Rosen},
  \citenamefont {Fox}, \citenamefont {Jiang}, \citenamefont {Lyu},
  \citenamefont {Li}, \citenamefont {Watanabe}, \citenamefont {Taniguchi},
  \citenamefont {Jung}, \citenamefont {Shi}, \citenamefont {Goldhaber-Gordon},
  \citenamefont {Zhang},\ and\ \citenamefont {Wang}}]{chen19n}%
  \BibitemOpen
  \bibfield  {author} {\bibinfo {author} {\bibfnamefont {G.}~\bibnamefont
  {Chen}}, \bibinfo {author} {\bibfnamefont {A.~L.}\ \bibnamefont {Sharpe}},
  \bibinfo {author} {\bibfnamefont {P.}~\bibnamefont {Gallagher}}, \bibinfo
  {author} {\bibfnamefont {I.~T.}\ \bibnamefont {Rosen}}, \bibinfo {author}
  {\bibfnamefont {E.~J.}\ \bibnamefont {Fox}}, \bibinfo {author} {\bibfnamefont
  {L.}~\bibnamefont {Jiang}}, \bibinfo {author} {\bibfnamefont
  {B.}~\bibnamefont {Lyu}}, \bibinfo {author} {\bibfnamefont {H.}~\bibnamefont
  {Li}}, \bibinfo {author} {\bibfnamefont {K.}~\bibnamefont {Watanabe}},
  \bibinfo {author} {\bibfnamefont {T.}~\bibnamefont {Taniguchi}}, \bibinfo
  {author} {\bibfnamefont {J.}~\bibnamefont {Jung}}, \bibinfo {author}
  {\bibfnamefont {Z.}~\bibnamefont {Shi}}, \bibinfo {author} {\bibfnamefont
  {D.}~\bibnamefont {Goldhaber-Gordon}}, \bibinfo {author} {\bibfnamefont
  {Y.}~\bibnamefont {Zhang}}, \ and\ \bibinfo {author} {\bibfnamefont
  {F.}~\bibnamefont {Wang}},\ }\bibfield  {title} {\enquote {\bibinfo {title}
  {Signatures of tunable superconductivity in a trilayer graphene moiré
  superlattice},}\ }\href {\doibase 10.1038/s41586-019-1393-y} {\bibfield
  {journal} {\bibinfo  {journal} {Nature}\ }\textbf {\bibinfo {volume} {572}},\
  \bibinfo {pages} {215} (\bibinfo {year} {2019})}\BibitemShut {NoStop}%
\bibitem [{\citenamefont {Bradlyn}\ \emph {et~al.}(2016)\citenamefont
  {Bradlyn}, \citenamefont {Cano}, \citenamefont {Wang}, \citenamefont
  {Vergniory}, \citenamefont {Felser}, \citenamefont {Cava},\ and\
  \citenamefont {Bernevig}}]{bradlyn16sc}%
  \BibitemOpen
  \bibfield  {author} {\bibinfo {author} {\bibfnamefont {B.}~\bibnamefont
  {Bradlyn}}, \bibinfo {author} {\bibfnamefont {J.}~\bibnamefont {Cano}},
  \bibinfo {author} {\bibfnamefont {Z.}~\bibnamefont {Wang}}, \bibinfo {author}
  {\bibfnamefont {M.~G.}\ \bibnamefont {Vergniory}}, \bibinfo {author}
  {\bibfnamefont {C.}~\bibnamefont {Felser}}, \bibinfo {author} {\bibfnamefont
  {R.~J.}\ \bibnamefont {Cava}}, \ and\ \bibinfo {author} {\bibfnamefont
  {B.~A.}\ \bibnamefont {Bernevig}},\ }\bibfield  {title} {\enquote {\bibinfo
  {title} {Beyond {Dirac} and {Weyl} fermions: Unconventional quasiparticles in
  conventional crystals},}\ }\href {\doibase 10.1126/science.aaf5037}
  {\bibfield  {journal} {\bibinfo  {journal} {Science}\ }\textbf {\bibinfo
  {volume} {353}},\ \bibinfo {pages} {aaf5037} (\bibinfo {year}
  {2016})}\BibitemShut {NoStop}%
\bibitem [{\citenamefont {Fulga}\ and\ \citenamefont
  {Stern}(2017)}]{fulga17prb}%
  \BibitemOpen
  \bibfield  {author} {\bibinfo {author} {\bibfnamefont {I.~C.}\ \bibnamefont
  {Fulga}}\ and\ \bibinfo {author} {\bibfnamefont {A.}~\bibnamefont {Stern}},\
  }\bibfield  {title} {\enquote {\bibinfo {title} {Triple point fermions in a
  minimal symmorphic model},}\ }\href {\doibase 10.1103/PhysRevB.95.241116}
  {\bibfield  {journal} {\bibinfo  {journal} {Phys. Rev. B}\ }\textbf {\bibinfo
  {volume} {95}},\ \bibinfo {pages} {241116} (\bibinfo {year}
  {2017})}\BibitemShut {NoStop}%
\bibitem [{\citenamefont {Zhu}\ \emph {et~al.}(2017)\citenamefont {Zhu},
  \citenamefont {Zhang}, \citenamefont {Yan}, \citenamefont {Xing},\ and\
  \citenamefont {Zhu}}]{zhu17pra}%
  \BibitemOpen
  \bibfield  {author} {\bibinfo {author} {\bibfnamefont {Y.-Q.}\ \bibnamefont
  {Zhu}}, \bibinfo {author} {\bibfnamefont {D.-W.}\ \bibnamefont {Zhang}},
  \bibinfo {author} {\bibfnamefont {H.}~\bibnamefont {Yan}}, \bibinfo {author}
  {\bibfnamefont {D.-Y.}\ \bibnamefont {Xing}}, \ and\ \bibinfo {author}
  {\bibfnamefont {S.-L.}\ \bibnamefont {Zhu}},\ }\bibfield  {title} {\enquote
  {\bibinfo {title} {Emergent pseudospin-1 {Maxwell} fermions with a threefold
  degeneracy in optical lattices},}\ }\href {\doibase
  10.1103/PhysRevA.96.033634} {\bibfield  {journal} {\bibinfo  {journal} {Phys.
  Rev. A}\ }\textbf {\bibinfo {volume} {96}},\ \bibinfo {pages} {033634}
  (\bibinfo {year} {2017})}\BibitemShut {NoStop}%
\bibitem [{\citenamefont {Hu}\ \emph {et~al.}(2018)\citenamefont {Hu},
  \citenamefont {Hou}, \citenamefont {Zhang},\ and\ \citenamefont
  {Zhang}}]{hu18prl}%
  \BibitemOpen
  \bibfield  {author} {\bibinfo {author} {\bibfnamefont {H.}~\bibnamefont
  {Hu}}, \bibinfo {author} {\bibfnamefont {J.}~\bibnamefont {Hou}}, \bibinfo
  {author} {\bibfnamefont {F.}~\bibnamefont {Zhang}}, \ and\ \bibinfo {author}
  {\bibfnamefont {C.}~\bibnamefont {Zhang}},\ }\bibfield  {title} {\enquote
  {\bibinfo {title} {Topological triply degenerate points induced by
  spin-tensor-momentum couplings},}\ }\href {\doibase
  10.1103/PhysRevLett.120.240401} {\bibfield  {journal} {\bibinfo  {journal}
  {Phys. Rev. Lett.}\ }\textbf {\bibinfo {volume} {120}},\ \bibinfo {pages}
  {240401} (\bibinfo {year} {2018})}\BibitemShut {NoStop}%
\bibitem [{\citenamefont {Ramires}\ and\ \citenamefont
  {Lado}(2019)}]{ramires19prb}%
  \BibitemOpen
  \bibfield  {author} {\bibinfo {author} {\bibfnamefont {A.}~\bibnamefont
  {Ramires}}\ and\ \bibinfo {author} {\bibfnamefont {J.~L.}\ \bibnamefont
  {Lado}},\ }\bibfield  {title} {\enquote {\bibinfo {title} {Impurity-induced
  triple point fermions in twisted bilayer graphene},}\ }\href {\doibase
  10.1103/PhysRevB.99.245118} {\bibfield  {journal} {\bibinfo  {journal} {Phys.
  Rev. B}\ }\textbf {\bibinfo {volume} {99}},\ \bibinfo {pages} {245118}
  (\bibinfo {year} {2019})}\BibitemShut {NoStop}%
\bibitem [{\citenamefont {Tan}\ \emph {et~al.}(2018)\citenamefont {Tan},
  \citenamefont {Zhang}, \citenamefont {Liu}, \citenamefont {Xue},
  \citenamefont {Yu}, \citenamefont {Zhu}, \citenamefont {Yan}, \citenamefont
  {Zhu},\ and\ \citenamefont {Yu}}]{tan18prl}%
  \BibitemOpen
  \bibfield  {author} {\bibinfo {author} {\bibfnamefont {X.}~\bibnamefont
  {Tan}}, \bibinfo {author} {\bibfnamefont {D.-W.}\ \bibnamefont {Zhang}},
  \bibinfo {author} {\bibfnamefont {Q.}~\bibnamefont {Liu}}, \bibinfo {author}
  {\bibfnamefont {G.}~\bibnamefont {Xue}}, \bibinfo {author} {\bibfnamefont
  {H.-F.}\ \bibnamefont {Yu}}, \bibinfo {author} {\bibfnamefont {Y.-Q.}\
  \bibnamefont {Zhu}}, \bibinfo {author} {\bibfnamefont {H.}~\bibnamefont
  {Yan}}, \bibinfo {author} {\bibfnamefont {S.-L.}\ \bibnamefont {Zhu}}, \ and\
  \bibinfo {author} {\bibfnamefont {Y.}~\bibnamefont {Yu}},\ }\bibfield
  {title} {\enquote {\bibinfo {title} {Topological {Maxwell} metal bands in a
  superconducting qutrit},}\ }\href {\doibase 10.1103/PhysRevLett.120.130503}
  {\bibfield  {journal} {\bibinfo  {journal} {Phys. Rev. Lett.}\ }\textbf
  {\bibinfo {volume} {120}},\ \bibinfo {pages} {130503} (\bibinfo {year}
  {2018})}\BibitemShut {NoStop}%
\bibitem [{\citenamefont {Venderbos}\ \emph {et~al.}(2016)\citenamefont
  {Venderbos}, \citenamefont {Kozii},\ and\ \citenamefont
  {Fu}}]{venderbos16prb}%
  \BibitemOpen
  \bibfield  {author} {\bibinfo {author} {\bibfnamefont {J.~W.~F.}\
  \bibnamefont {Venderbos}}, \bibinfo {author} {\bibfnamefont {V.}~\bibnamefont
  {Kozii}}, \ and\ \bibinfo {author} {\bibfnamefont {L.}~\bibnamefont {Fu}},\
  }\bibfield  {title} {\enquote {\bibinfo {title} {Odd-parity superconductors
  with two-component order parameters: Nematic and chiral, full gap, and
  {Majorana} node},}\ }\href {\doibase 10.1103/PhysRevB.94.180504} {\bibfield
  {journal} {\bibinfo  {journal} {Phys. Rev. B}\ }\textbf {\bibinfo {volume}
  {94}},\ \bibinfo {pages} {180504} (\bibinfo {year} {2016})}\BibitemShut
  {NoStop}%
\bibitem [{\citenamefont {Kozii}\ \emph {et~al.}(2016)\citenamefont {Kozii},
  \citenamefont {Venderbos},\ and\ \citenamefont {Fu}}]{kozii16sa}%
  \BibitemOpen
  \bibfield  {author} {\bibinfo {author} {\bibfnamefont {V.}~\bibnamefont
  {Kozii}}, \bibinfo {author} {\bibfnamefont {J.~W.~F.}\ \bibnamefont
  {Venderbos}}, \ and\ \bibinfo {author} {\bibfnamefont {L.}~\bibnamefont
  {Fu}},\ }\bibfield  {title} {\enquote {\bibinfo {title} {Three-dimensional
  {Majorana} fermions in chiral superconductors},}\ }\href {\doibase
  10.1126/sciadv.1601835} {\bibfield  {journal} {\bibinfo  {journal} {Sci.
  Adv.}\ }\textbf {\bibinfo {volume} {2}},\ \bibinfo {pages} {e1601835}
  (\bibinfo {year} {2016})}\BibitemShut {NoStop}%
\bibitem [{\citenamefont {Meng}\ and\ \citenamefont
  {Balents}(2012)}]{meng12prb}%
  \BibitemOpen
  \bibfield  {author} {\bibinfo {author} {\bibfnamefont {T.}~\bibnamefont
  {Meng}}\ and\ \bibinfo {author} {\bibfnamefont {L.}~\bibnamefont {Balents}},\
  }\bibfield  {title} {\enquote {\bibinfo {title} {Weyl superconductors},}\
  }\href {\doibase 10.1103/PhysRevB.86.054504} {\bibfield  {journal} {\bibinfo
  {journal} {Phys. Rev. B}\ }\textbf {\bibinfo {volume} {86}},\ \bibinfo
  {pages} {054504} (\bibinfo {year} {2012})}\BibitemShut {NoStop}%
\bibitem [{\citenamefont {Sau}\ and\ \citenamefont {Tewari}(2012)}]{sau12prb}%
  \BibitemOpen
  \bibfield  {author} {\bibinfo {author} {\bibfnamefont {J.~D.}\ \bibnamefont
  {Sau}}\ and\ \bibinfo {author} {\bibfnamefont {S.}~\bibnamefont {Tewari}},\
  }\bibfield  {title} {\enquote {\bibinfo {title} {Topologically protected
  surface {Majorana} arcs and bulk {Weyl} fermions in ferromagnetic
  superconductors},}\ }\href {\doibase 10.1103/PhysRevB.86.104509} {\bibfield
  {journal} {\bibinfo  {journal} {Phys. Rev. B}\ }\textbf {\bibinfo {volume}
  {86}},\ \bibinfo {pages} {104509} (\bibinfo {year} {2012})}\BibitemShut
  {NoStop}%
\bibitem [{\citenamefont {Wu}\ and\ \citenamefont {Yang}(1976)}]{wu76npb}%
  \BibitemOpen
  \bibfield  {author} {\bibinfo {author} {\bibfnamefont {T.~T.}\ \bibnamefont
  {Wu}}\ and\ \bibinfo {author} {\bibfnamefont {C.~N.}\ \bibnamefont {Yang}},\
  }\bibfield  {title} {\enquote {\bibinfo {title} {Dirac monopole without
  strings: Monopole harmonics},}\ }\href {\doibase
  https://doi.org/10.1016/0550-3213(76)90143-7} {\bibfield  {journal} {\bibinfo
   {journal} {Nucl. Phys B}\ }\textbf {\bibinfo {volume} {107}},\ \bibinfo
  {pages} {365 } (\bibinfo {year} {1976})}\BibitemShut {NoStop}%
\bibitem [{\citenamefont {Wu}\ and\ \citenamefont {Yang}(1977)}]{wu77prd}%
  \BibitemOpen
  \bibfield  {author} {\bibinfo {author} {\bibfnamefont {T.~T.}\ \bibnamefont
  {Wu}}\ and\ \bibinfo {author} {\bibfnamefont {C.~N.}\ \bibnamefont {Yang}},\
  }\bibfield  {title} {\enquote {\bibinfo {title} {Some properties of monopole
  harmonics},}\ }\href {\doibase 10.1103/PhysRevD.16.1018} {\bibfield
  {journal} {\bibinfo  {journal} {Phys. Rev. D}\ }\textbf {\bibinfo {volume}
  {16}},\ \bibinfo {pages} {1018} (\bibinfo {year} {1977})}\BibitemShut
  {NoStop}%
\bibitem [{\citenamefont {Xiao}\ \emph {et~al.}(2010)\citenamefont {Xiao},
  \citenamefont {Chang},\ and\ \citenamefont {Niu}}]{xiao10rmp}%
  \BibitemOpen
  \bibfield  {author} {\bibinfo {author} {\bibfnamefont {D.}~\bibnamefont
  {Xiao}}, \bibinfo {author} {\bibfnamefont {M.-C.}\ \bibnamefont {Chang}}, \
  and\ \bibinfo {author} {\bibfnamefont {Q.}~\bibnamefont {Niu}},\ }\bibfield
  {title} {\enquote {\bibinfo {title} {Berry phase effects on electronic
  properties},}\ }\href {\doibase 10.1103/RevModPhys.82.1959} {\bibfield
  {journal} {\bibinfo  {journal} {Rev. Mod. Phys.}\ }\textbf {\bibinfo {volume}
  {82}},\ \bibinfo {pages} {1959} (\bibinfo {year} {2010})}\BibitemShut
  {NoStop}%
\bibitem [{\citenamefont {Cho}\ \emph {et~al.}(2012)\citenamefont {Cho},
  \citenamefont {Bardarson}, \citenamefont {Lu},\ and\ \citenamefont
  {Moore}}]{cho12prb}%
  \BibitemOpen
  \bibfield  {author} {\bibinfo {author} {\bibfnamefont {G.~Y.}\ \bibnamefont
  {Cho}}, \bibinfo {author} {\bibfnamefont {J.~H.}\ \bibnamefont {Bardarson}},
  \bibinfo {author} {\bibfnamefont {Y.-M.}\ \bibnamefont {Lu}}, \ and\ \bibinfo
  {author} {\bibfnamefont {J.~E.}\ \bibnamefont {Moore}},\ }\bibfield  {title}
  {\enquote {\bibinfo {title} {Superconductivity of doped {Weyl} semimetals:
  Finite-momentum pairing and electronic analog of the {${}^{3}$He-$A$}
  phase},}\ }\href {\doibase 10.1103/PhysRevB.86.214514} {\bibfield  {journal}
  {\bibinfo  {journal} {Phys. Rev. B}\ }\textbf {\bibinfo {volume} {86}},\
  \bibinfo {pages} {214514} (\bibinfo {year} {2012})}\BibitemShut {NoStop}%
\bibitem [{\citenamefont {Savary}\ \emph {et~al.}(2017)\citenamefont {Savary},
  \citenamefont {Ruhman}, \citenamefont {Venderbos}, \citenamefont {Fu},\ and\
  \citenamefont {Lee}}]{savary17prb}%
  \BibitemOpen
  \bibfield  {author} {\bibinfo {author} {\bibfnamefont {L.}~\bibnamefont
  {Savary}}, \bibinfo {author} {\bibfnamefont {J.}~\bibnamefont {Ruhman}},
  \bibinfo {author} {\bibfnamefont {J.~W.~F.}\ \bibnamefont {Venderbos}},
  \bibinfo {author} {\bibfnamefont {L.}~\bibnamefont {Fu}}, \ and\ \bibinfo
  {author} {\bibfnamefont {P.~A.}\ \bibnamefont {Lee}},\ }\bibfield  {title}
  {\enquote {\bibinfo {title} {Superconductivity in three-dimensional
  spin-orbit coupled semimetals},}\ }\href {\doibase
  10.1103/PhysRevB.96.214514} {\bibfield  {journal} {\bibinfo  {journal} {Phys.
  Rev. B}\ }\textbf {\bibinfo {volume} {96}},\ \bibinfo {pages} {214514}
  (\bibinfo {year} {2017})}\BibitemShut {NoStop}%
\bibitem [{\citenamefont {Agterberg}\ \emph {et~al.}(2020)\citenamefont
  {Agterberg}, \citenamefont {Davis}, \citenamefont {Edkins}, \citenamefont
  {Fradkin}, \citenamefont {Van~Harlingen}, \citenamefont {Kivelson},
  \citenamefont {Lee}, \citenamefont {Radzihovsky}, \citenamefont {Tranquada},\
  and\ \citenamefont {Wang}}]{agterberg20ar}%
  \BibitemOpen
  \bibfield  {author} {\bibinfo {author} {\bibfnamefont {D.~F.}\ \bibnamefont
  {Agterberg}}, \bibinfo {author} {\bibfnamefont {J.~S.}\ \bibnamefont
  {Davis}}, \bibinfo {author} {\bibfnamefont {S.~D.}\ \bibnamefont {Edkins}},
  \bibinfo {author} {\bibfnamefont {E.}~\bibnamefont {Fradkin}}, \bibinfo
  {author} {\bibfnamefont {D.~J.}\ \bibnamefont {Van~Harlingen}}, \bibinfo
  {author} {\bibfnamefont {S.~A.}\ \bibnamefont {Kivelson}}, \bibinfo {author}
  {\bibfnamefont {P.~A.}\ \bibnamefont {Lee}}, \bibinfo {author} {\bibfnamefont
  {L.}~\bibnamefont {Radzihovsky}}, \bibinfo {author} {\bibfnamefont {J.~M.}\
  \bibnamefont {Tranquada}}, \ and\ \bibinfo {author} {\bibfnamefont
  {Y.}~\bibnamefont {Wang}},\ }\bibfield  {title} {\enquote {\bibinfo {title}
  {The physics of pair-density waves: Cuprate superconductors and beyond},}\
  }\href {\doibase 10.1146/annurev-conmatphys-031119-050711} {\bibfield
  {journal} {\bibinfo  {journal} {Annu. Rev. Cond. Matter Phys.}\ }\textbf
  {\bibinfo {volume} {11}},\ \bibinfo {pages} {231} (\bibinfo {year}
  {2020})}\BibitemShut {NoStop}%
\bibitem [{\citenamefont {Tinkham}(2004)}]{tinkhambook}%
  \BibitemOpen
  \bibfield  {author} {\bibinfo {author} {\bibfnamefont {M.}~\bibnamefont
  {Tinkham}},\ }\href@noop {} {\emph {\bibinfo {title} {Introduction to
  Superconductivity}}}\ (\bibinfo  {publisher} {Dover Publication, New York},\
  \bibinfo {year} {2004})\BibitemShut {NoStop}%
\bibitem [{\citenamefont {Kosterlitz}\ and\ \citenamefont
  {Thouless}(1973)}]{kosterlitz73jpc}%
  \BibitemOpen
  \bibfield  {author} {\bibinfo {author} {\bibfnamefont {J.~M.}\ \bibnamefont
  {Kosterlitz}}\ and\ \bibinfo {author} {\bibfnamefont {D.~J.}\ \bibnamefont
  {Thouless}},\ }\bibfield  {title} {\enquote {\bibinfo {title} {Ordering,
  metastability and phase transitions in two-dimensional systems},}\ }\href
  {\doibase 10.1088/0022-3719/6/7/010} {\bibfield  {journal} {\bibinfo
  {journal} {J. Phys. C}\ }\textbf {\bibinfo {volume} {6}},\ \bibinfo {pages}
  {1181} (\bibinfo {year} {1973})}\BibitemShut {NoStop}%
\bibitem [{\citenamefont {Hazra}\ \emph {et~al.}(2019)\citenamefont {Hazra},
  \citenamefont {Verma},\ and\ \citenamefont {Randeria}}]{hazra19prx}%
  \BibitemOpen
  \bibfield  {author} {\bibinfo {author} {\bibfnamefont {T.}~\bibnamefont
  {Hazra}}, \bibinfo {author} {\bibfnamefont {N.}~\bibnamefont {Verma}}, \ and\
  \bibinfo {author} {\bibfnamefont {M.}~\bibnamefont {Randeria}},\ }\bibfield
  {title} {\enquote {\bibinfo {title} {Bounds on the superconducting transition
  temperature: Applications to twisted bilayer graphene and cold atoms},}\
  }\href {\doibase 10.1103/PhysRevX.9.031049} {\bibfield  {journal} {\bibinfo
  {journal} {Phys. Rev. X}\ }\textbf {\bibinfo {volume} {9}},\ \bibinfo {pages}
  {031049} (\bibinfo {year} {2019})}\BibitemShut {NoStop}%
\bibitem [{\citenamefont {Emery}\ and\ \citenamefont
  {Kivelson}(1995)}]{emery95n}%
  \BibitemOpen
  \bibfield  {author} {\bibinfo {author} {\bibfnamefont {V.~J.}\ \bibnamefont
  {Emery}}\ and\ \bibinfo {author} {\bibfnamefont {S.~A.}\ \bibnamefont
  {Kivelson}},\ }\bibfield  {title} {\enquote {\bibinfo {title} {Importance of
  phase fluctuations in superconductors with small superfluid density},}\
  }\href {\doibase 10.1038/374434a0} {\bibfield  {journal} {\bibinfo  {journal}
  {Nature}\ }\textbf {\bibinfo {volume} {374}},\ \bibinfo {pages} {434}
  (\bibinfo {year} {1995})}\BibitemShut {NoStop}%
\bibitem [{\citenamefont {Engelbrecht}\ \emph {et~al.}(1997)\citenamefont
  {Engelbrecht}, \citenamefont {Randeria},\ and\ \citenamefont
  {S\'ade~Melo}}]{engelbrecht97prb}%
  \BibitemOpen
  \bibfield  {author} {\bibinfo {author} {\bibfnamefont {J.~R.}\ \bibnamefont
  {Engelbrecht}}, \bibinfo {author} {\bibfnamefont {M.}~\bibnamefont
  {Randeria}}, \ and\ \bibinfo {author} {\bibfnamefont {C.~A.~R.}\ \bibnamefont
  {S\'ade~Melo}},\ }\bibfield  {title} {\enquote {\bibinfo {title} {{BCS} to
  {Bose} crossover: Broken-symmetry state},}\ }\href {\doibase
  10.1103/PhysRevB.55.15153} {\bibfield  {journal} {\bibinfo  {journal} {Phys.
  Rev. B}\ }\textbf {\bibinfo {volume} {55}},\ \bibinfo {pages} {15153}
  (\bibinfo {year} {1997})}\BibitemShut {NoStop}%
\bibitem [{\citenamefont {Altland}\ and\ \citenamefont
  {Simons}(2010)}]{altlandbook}%
  \BibitemOpen
  \bibfield  {author} {\bibinfo {author} {\bibfnamefont {A.}~\bibnamefont
  {Altland}}\ and\ \bibinfo {author} {\bibfnamefont {B.~D.}\ \bibnamefont
  {Simons}},\ }\href@noop {} {\emph {\bibinfo {title} {Condensed Matter Field
  Theory}}},\ \bibinfo {edition} {2nd}\ ed.\ (\bibinfo  {publisher} {Cambridge
  University Press, Cambridge},\ \bibinfo {year} {2010})\BibitemShut {NoStop}%
\bibitem [{\citenamefont {Elliott}\ and\ \citenamefont
  {Franz}(2015)}]{elliott15rmp}%
  \BibitemOpen
  \bibfield  {author} {\bibinfo {author} {\bibfnamefont {S.~R.}\ \bibnamefont
  {Elliott}}\ and\ \bibinfo {author} {\bibfnamefont {M.}~\bibnamefont
  {Franz}},\ }\bibfield  {title} {\enquote {\bibinfo {title} {Colloquium:
  Majorana fermions in nuclear, particle, and solid-state physics},}\ }\href
  {\doibase 10.1103/RevModPhys.87.137} {\bibfield  {journal} {\bibinfo
  {journal} {Rev. Mod. Phys.}\ }\textbf {\bibinfo {volume} {87}},\ \bibinfo
  {pages} {137} (\bibinfo {year} {2015})}\BibitemShut {NoStop}%
\bibitem [{\citenamefont {Armitage}\ \emph {et~al.}(2018)\citenamefont
  {Armitage}, \citenamefont {Mele},\ and\ \citenamefont
  {Vishwanath}}]{armitage18rmp}%
  \BibitemOpen
  \bibfield  {author} {\bibinfo {author} {\bibfnamefont {N.~P.}\ \bibnamefont
  {Armitage}}, \bibinfo {author} {\bibfnamefont {E.~J.}\ \bibnamefont {Mele}},
  \ and\ \bibinfo {author} {\bibfnamefont {A.}~\bibnamefont {Vishwanath}},\
  }\bibfield  {title} {\enquote {\bibinfo {title} {Weyl and {Dirac} semimetals
  in three-dimensional solids},}\ }\href {\doibase
  10.1103/RevModPhys.90.015001} {\bibfield  {journal} {\bibinfo  {journal}
  {Rev. Mod. Phys.}\ }\textbf {\bibinfo {volume} {90}},\ \bibinfo {pages}
  {015001} (\bibinfo {year} {2018})}\BibitemShut {NoStop}%
\bibitem [{\citenamefont {Li}\ and\ \citenamefont {Haldane}(2018)}]{li18prl}%
  \BibitemOpen
  \bibfield  {author} {\bibinfo {author} {\bibfnamefont {Y.}~\bibnamefont
  {Li}}\ and\ \bibinfo {author} {\bibfnamefont {F.~D.~M.}\ \bibnamefont
  {Haldane}},\ }\bibfield  {title} {\enquote {\bibinfo {title} {Topological
  nodal {Cooper} pairing in doped {Weyl} metals},}\ }\href {\doibase
  10.1103/PhysRevLett.120.067003} {\bibfield  {journal} {\bibinfo  {journal}
  {Phys. Rev. Lett.}\ }\textbf {\bibinfo {volume} {120}},\ \bibinfo {pages}
  {067003} (\bibinfo {year} {2018})}\BibitemShut {NoStop}%
\bibitem [{\citenamefont {Marsiglio}(2020)}]{marsiglio20ap}%
  \BibitemOpen
  \bibfield  {author} {\bibinfo {author} {\bibfnamefont {F.}~\bibnamefont
  {Marsiglio}},\ }\bibfield  {title} {\enquote {\bibinfo {title} {Eliashberg
  theory: A short review},}\ }\href {\doibase
  https://doi.org/10.1016/j.aop.2020.168102} {\bibfield  {journal} {\bibinfo
  {journal} {Ann. Phys.}\ }\textbf {\bibinfo {volume} {417}},\ \bibinfo {pages}
  {168102} (\bibinfo {year} {2020})}\BibitemShut {NoStop}%
\bibitem [{\citenamefont {Wang}\ and\ \citenamefont {Ye}(2016)}]{wangye16prb}%
  \BibitemOpen
  \bibfield  {author} {\bibinfo {author} {\bibfnamefont {Y.}~\bibnamefont
  {Wang}}\ and\ \bibinfo {author} {\bibfnamefont {P.}~\bibnamefont {Ye}},\
  }\bibfield  {title} {\enquote {\bibinfo {title} {Topological density-wave
  states in a particle-hole symmetric {Weyl} metal},}\ }\href {\doibase
  10.1103/PhysRevB.94.075115} {\bibfield  {journal} {\bibinfo  {journal} {Phys.
  Rev. B}\ }\textbf {\bibinfo {volume} {94}},\ \bibinfo {pages} {075115}
  (\bibinfo {year} {2016})}\BibitemShut {NoStop}%
\bibitem [{\citenamefont {{Sim}}\ \emph {et~al.}(2019)\citenamefont {{Sim}},
  \citenamefont {{Park}},\ and\ \citenamefont {{Lee}}}]{sim19ax}%
  \BibitemOpen
  \bibfield  {author} {\bibinfo {author} {\bibfnamefont {G.}~\bibnamefont
  {{Sim}}}, \bibinfo {author} {\bibfnamefont {M.~J.}\ \bibnamefont {{Park}}}, \
  and\ \bibinfo {author} {\bibfnamefont {S.}~\bibnamefont {{Lee}}},\ }\bibfield
   {title} {\enquote {\bibinfo {title} {{Triplet-Superconductivity in
  Triple-Band Crossings}},}\ }\href@noop {} {\bibfield  {journal} {\bibinfo
  {journal} {arXiv e-prints}\ ,\ \bibinfo {eid} {arXiv:1909.04015}} (\bibinfo
  {year} {2019})},\ \Eprint {http://arxiv.org/abs/1909.04015} {arXiv:1909.04015
  [cond-mat.str-el]} \BibitemShut {NoStop}%
\end{thebibliography}%

\end{document}